\PassOptionsToPackage{table}{xcolor}
\documentclass[conference]{IEEEtran}
\usepackage{multirow}
\usepackage{todonotes}
\usepackage{regexpatch}
\makeatletter
\xpatchcmd{\@todo}{\setkeys{todonotes}{#1}}{\setkeys{todonotes}{inline,#1}}{}{}
\makeatother
\usepackage{makecell}
\usepackage{amsmath}
\usepackage{amsfonts}
\usepackage{amsthm}
\usepackage{enumerate}
\usepackage{xspace}
\newtheorem{definition}{Definition}
\usepackage{colortbl}%for color table cells
\usepackage{nicematrix}
\usepackage{flushend}

\def \FF {{\mathbb F}}
\newcommand{\dis}[1]{\textrm{dis}(#1)}

\newcommand{\later}[1]{\textcolor{black}{#1}}%date to be updated later when the experiments are finished

\definecolor{color1}{HTML}{176BA0}
\definecolor{color2}{HTML}{C7F9EE}
\definecolor{color3}{HTML}{142459}
\def \coften {\cellcolor{color1!35}}
\def \rowcone {white}
\def \rowctwo {gray!35}

\newcommand{\method}{DeepNcode\xspace}
\begin{document}

\title{DeepNcode: Encoding-Based Protection against Bit-Flip Attacks on Neural Networks}

\author{Patrik Vel\v{c}ick\'y$^1$, Jakub Breier$^2$, Mladen Kova\v{c}evi\'c$^{3,4}$, Xiaolu Hou$^1$\\
$^1$Faculty of Informatics and Information Technologies, Slovak University of Technology, Bratislava, Slovakia\\
\texttt{xvelcickyp@fiit.stuba.sk, houxiaolu.email@gmail.com}\\
$^2$TTControl GmbH, Vienna, Austria\\
\texttt{jbreier@jbreier.com}\\
$^3$Faculty of Technical Sciences, University of Novi Sad, Serbia\\
$^4$Ethernal, Novi Sad, Serbia\\
\texttt{kmladen@uns.ac.rs}}

% \IEEEoverridecommandlockouts
% \makeatletter\def\@IEEEpubidpullup{6.5\baselineskip}\makeatother
% \IEEEpubid{\parbox{\columnwidth}{
%     Network and Distributed System Security (NDSS) Symposium 2024\\
%     26 February - 1 March 2024, San Diego, CA, USA\\
%     ISBN 1-891562-93-2\\
%     https://dx.doi.org/10.14722/ndss.2024.23xxx\\
%     www.ndss-symposium.org
% }
% \hspace{\columnsep}\makebox[\columnwidth]{}}

% \affiliation{%

% \affiliation{%

\maketitle

\begin{abstract}
Fault injection attacks are a potent threat against embedded implementations of neural network models.
Several attack vectors have been proposed, such as misclassification, model extraction, and trojan/backdoor planting.
Most of these attacks work by flipping bits in the memory where quantized model parameters are stored.

In this paper, we introduce an encoding-based protection method against bit-flip attacks on neural networks, titled DeepNcode.
We experimentally evaluate our proposal with several publicly available models and datasets, by using state-of-the-art bit-flip attacks: BFA, T-BFA, and TA-LBF.
Our results show an increase in protection margin of up to $7.6\times$ for $4-$bit and $12.4\times$ for $8-$bit quantized networks.
Memory overheads start at $50\%$ of the original network size, while the time overheads are negligible.
Moreover, DeepNcode does not require retraining and does not change the original accuracy of the model.
\end{abstract}


% \keywords{neural networks, hardware security, fault injection attacks, countermeasures, coding theory}




\section{Introduction}
Deep learning has become a workhorse for many applications that can be successfully accomplished by means of statistical learning.
This includes security- and safety-critical tasks, such as healthcare, autonomous driving, network security, connected mobile machinery, etc.
With the widespread adoption of deep neural networks (DNNs), these models are now being used in embedded devices, often in the form of Internet-of-Things (IoT) appliances, that can be deployed anywhere, within the potential physical reach of an attacker.

The reliance on DNNs poses a significant threat to the security and privacy of sensitive information.
Initially, researchers pointed out that imperceptible input perturbations can cause a DNN model to misclassify~\cite{szegedy2013intriguing}.
This property immediately caught the attention of the computer security community and paved the way to the newly established area -- adversarial attacks on machine learning.
Many attack vectors have emerged, ranging from a simple misclassification to poisoning the training dataset, to inferring the information about the training data~\cite{biggio2018wild}.
In the case of embedded applications, hardware security vectors can cause security breaches that are normally not considered within the adversarial learning scope which assumes either input perturbations or tampering with the learning process~\cite{batina2022implementation,mittal2021survey}.
It was shown that fault injection attack is a realistic threat against embedded DNN models~\cite{breier2018practical,hong2019terminal,mittal2021survey}.
Several attack scenarios have been proposed up to date that successfully utilize fault attacks, such as misclassification~\cite{liu2017fault}, reverse engineering~\cite{breier2021sniff}, trojan planting~\cite{rakin2020tbt}, and backdoor planting during the training~\cite{breier2022foobar}.
It is therefore natural that researchers started developing protection techniques against this powerful attack~\cite{rakin2021ra,he2020defending}.
As with any countermeasure method, there are trade-offs between security and efficiency, and the implementer's goal is to find the sweet spot that adequately matches the required security level but does not incur overwhelming overhead.
There is no one-size-fits-all solution, as some of the attacks focus on disturbing weights (mostly via the Rowhammer attack model~\cite{kim2014flipping}), and others on tampering with the execution routine (utilizing, for example, instruction skips~\cite{breier2018practical}).
Similarly, protecting GPU-deployed models is different from hardening models on embedded devices.

\begin{figure}
    \centering
    \setlength{\tabcolsep}{1pt}
    \begin{tabular}{cc}
        \includegraphics[width=0.2325\textwidth]{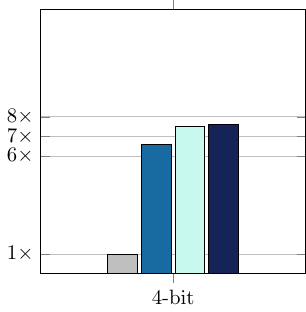} & \includegraphics[width=0.24\textwidth]{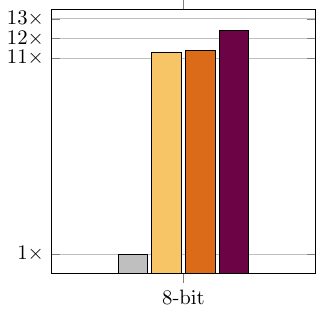}\\
        \multicolumn{2}{c}{\hspace{5mm}\includegraphics[trim={0 0 0 5.8cm},clip, width=0.4\textwidth]{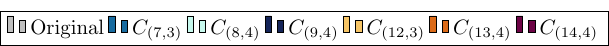}}
    \end{tabular}
    
    \caption{Summary of the improvement in the number of bit flips required for the successful attack with \method over the unprotected implementation for 4-bit and 8-bit quantized networks. The data is the average of the minimum number of bit flips over all the networks and datasets analyzed in this work. Detailed results can be found in tables~\ref{tab:4-bit-result} and~\ref{tab:8-bit-results}.}
    \label{fig:overview}
\end{figure}

\begin{figure}
    \centering
    \includegraphics[width=0.48\textwidth]{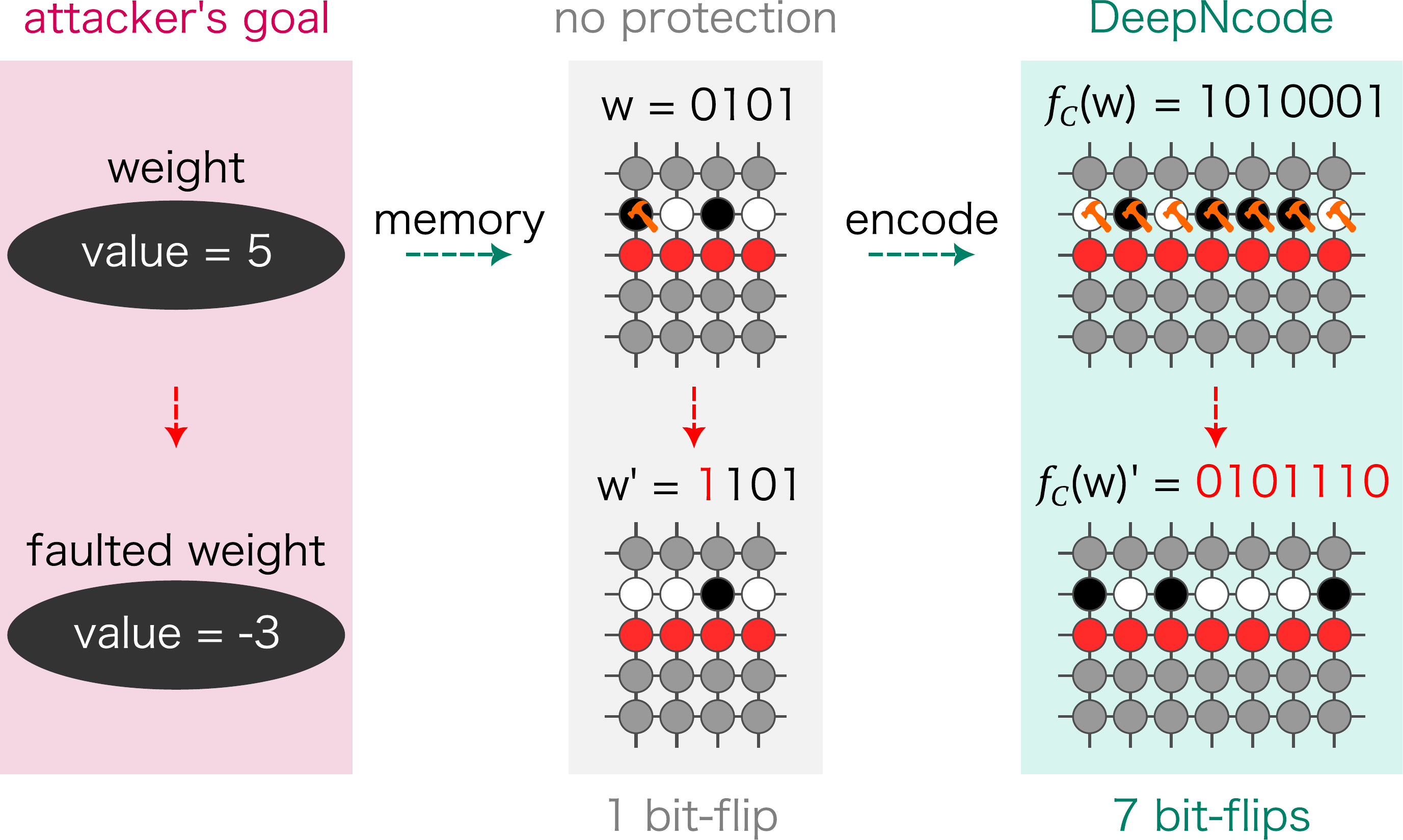}
    \caption{High-level overview of \method. The attacker's goal is to change the original weight value (e.g., $5$) to another value (e.g., $-3$). To do that, a Rowhammer can be used to flip bits in memory.
    In this particular example, while in the traditional two's complement representation, this can be done by a single bit flip, with the \method protection, all the $7$ bits need to be flipped, thus drastically increasing the attacker's effort.} 
    \label{fig:deepncode}
\end{figure}

In this paper, for the first time, we explore the potential of binary codes to significantly increase the attacker's effort in terms of injected bit flips.
By utilizing these codes, we also get formally proven security guarantees that stem from the code properties.
As shown in Figure~\ref{fig:overview}, the security margin in terms of number of bit flips required for a successful attack is increased $6.6 - 7.6\times$ for models with $4-$bit quantization and $11.3 - 12.4\times$ for models with $8-$bit quantization.
A high-level overview of our approach, titled \method, is depicted in Figure~\ref{fig:deepncode}.
In bit-flip attacks, the attacker aims to change the quantized weight values, which are stored in the memory.
Without the protection, these are stored in two's complement representation, allowing the attacker to do the majority of weight changes with just a single bit-flip.
By utilizing binary codes in \method, the attacker's effort is greatly increased, in this particular case by $7\times$.

\noindent
Our technical contributions can be summarized as follows:
\begin{itemize}
    \item We present \method countermeasure which tailors the code properties based on the weaknesses of the commonly used two's complement integer representation, standardly used for quantized neural networks.
    These weaknesses have been successfully exploited by bit-flip attacks that aim to minimize number of faults to cause as much damage as possible in terms of accuracy degradation.
    \item We experimentally evaluate \method on publicly available models (ResNet-20, ResNet-18, VGG-11) and datasets (CIFAR-10, CIFAR-100), using state-of-the-art attack methods, such as BFA~\cite{rakin2019bit}, T-BFA~\cite{rakin2021t}, and TA-LBF~\cite{bai2021targeted}.
    Our results show high security margins against these attacks.
    \item \method does not need retraining of the model and does not change the accuracy of the network. 
    It can be generally used on any $4-$bit or $8-$bit quantized network, irrespective of the number of parameters or architecture.
    \item As encoding methods provide provable security guarantees based on the underlying code parameters, the number of required bit flips for a successful attack will always be several times higher compared to attacking unprotected model, even if new more efficient attack is found.
\end{itemize}

\section{Background}
In this section, we introduce the main concepts necessary to understand the presented methods.
First, we provide a brief overview of neural networks and quantized neural networks.
Then, we detail the state-of-the-art fault attacks on neural networks.
Finally, we give overview of the protection methods against fault attacks.

\begin{figure*}[t]
    \centering
    \includegraphics[width=1.0\textwidth]{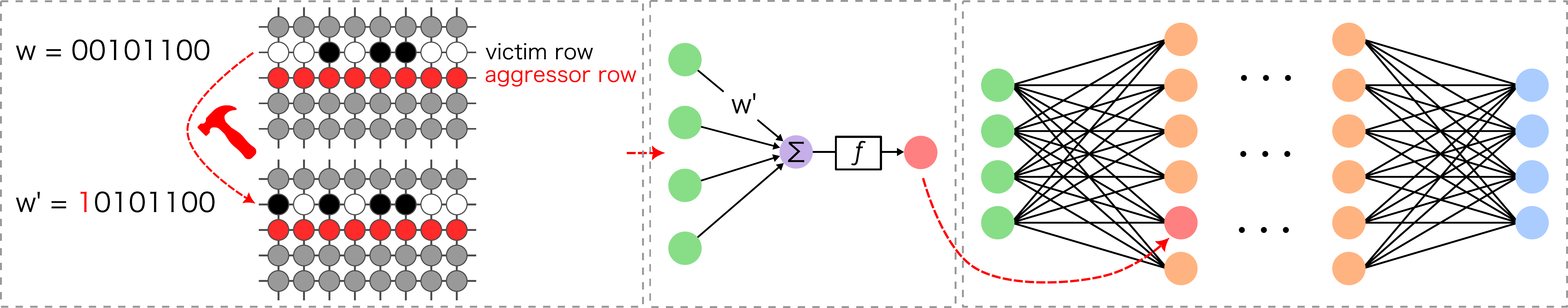}
    \caption{Illustration of a bit-flip attack on neural networks by utilizing the Rowhammer technique. The attacker flips a bit in the DDR memory by rapidly writing into the aggressor row. This changes the weight value stored in the victim row. During the inference phase, this alters the output value of the target neuron, and if enough neurons are hammered, the output class is changed.}
    \label{fig:rowhammer}
\end{figure*}

\subsection{Neural Networks}
Neural networks (NNs) are a type of machine learning algorithm inspired by the structure and function of the human brain~\cite{lecun2015deep}. 
They have gained tremendous popularity in recent years due to their ability to solve complex problems in various domains such as image and speech recognition, natural language processing, anomaly detection, and autonomous control systems.

At their core, NNs are composed of interconnected nodes, called \textit{neurons}, that process and transmit information through weighted connections. 
Through a process called training, NNs can learn to identify patterns in data and make accurate predictions or classifications.

NNs are highly versatile and can be customized to suit various applications, from simple linear regression models to deep neural networks with hundreds of layers. 
The use of NNs has led to significant advances in fields such as computer vision, natural language processing, and robotics, and continues to drive innovation in many areas of research and industry.

\noindent
\textbf{Quantized neural networks.}
Quantized neural networks (QNNs) have emerged as a promising approach for reducing the computational complexity and memory requirements of traditional neural networks while maintaining high accuracy. 
QNNs use low-precision data representations, typically in the form of integers, to perform computations and store weights, leading to significant reductions in memory usage and energy consumption~\cite{hubara2017quantized}.

QNNs have gained attention in recent years due to the growing demand for efficient and low-power neural network implementations in resource-constrained environments, such as mobile devices, IoT devices, and embedded systems. 
Moreover, the use of QNNs can reduce the cost of hardware design and enable the deployment of NNs on custom hardware with limited resources.
That is why QNNs are often the primary choice for embedded and IoT devices that have constrained computational power and memory.

In QNNs, the quantization process introduces non-linearity and discretization noise, which can affect the accuracy of the model. 
To mitigate these effects, various quantization techniques have been proposed, including uniform and non-uniform quantization, symmetric and asymmetric quantization, and mixed-precision quantization~\cite{nagel2021white}.

\subsection{Fault Attacks Against Neural Networks}
Fault attacks can take various forms, such as voltage glitches, laser fault injection, and electromagnetic interference, among others~\cite{breier2022practical}. 
In the area of cryptography, these attacks have been well-studied since their inception in 1997 when it was shown that a single fault can reveal the private key of the RSA cryptosystem~\cite{boneh2001importance}.
Generally, every symmetric cryptosystem is vulnerable to fault attacks, if the implementation is not specifically protected against them~\cite{baksi2022survey}.

In the area of neural networks, the first work utilizing fault attacks was published by Liu et al. in 2017~\cite{liu2017fault}, where the authors proposed bit flipping to induce misclassification.
This was followed by the first practical laser fault injection by Breier et al.~\cite{breier2018practical}, showing misclassification caused by instruction skips.
Later, this physical technique was shown to be effective with regards to other attack goals, such as model extraction~\cite{breier2021sniff}, trojan insertion~\cite{rakin2020tbt}, and backdoor injection~\cite{breier2022foobar}.
A summary of the state-of-the-art fault attacks on neural networks is given in Table~\ref{tab:fia_dl}.

When it comes to fault attacks on quantized neural networks, in most of the recently published works, the fault model is a single-bit flip, normally on multiple weight values stored in the memory.
Such an attack can be realized, for example, by utilizing the Rowhammer bug in computer memories~\cite{mutlu2019rowhammer}.
The proposed \method countermeasure aims to protect against such attacks.
It applies to, e.g., bit-flip attack (BFA)~\cite{rakin2019bit}, targeted bit-flip attack (T-BFA)~\cite{rakin2021t}, and targeted attack with limited bit-flips (TA-LBF)~\cite{bai2021targeted}, to name just a few of the recent works in this field.
The idea of a bit-flip-based attack is depicted in Figure~\ref{fig:rowhammer}.
The attacker aims at changing the model weights to corrupt the model behavior -- most commonly, the result would be an input misclassification.
To do this, an aggressive reading/writing to the memory cells (``aggressor row'') adjacent to the ones holding the target weight (``victim row'') is executed.
As memories nowadays exhibit high cell density to fit larger information capacity within a small space, this produces unwanted electromagnetic interactions, ultimately leading to data changes in the victim row.

\begin{table}[tb]
\begin{center}
 \begin{tabular}{|c|c|c|c|} 
 \hline
 Work & Technique & Attack Goal & Network \\ %[0.5ex] 
 \hline\hline
 \cite{liu2017fault} & bit flip & misclassification & MLP  \\\hline
 \cite{breier2018practical},\cite{hou2021physical} & instruction skip & misclassification & MLP \\\hline
\multirow{3}{*}{\cite{hong2019terminal}} & \multirow{3}{*}{bit flip} & \multirow{3}{*}{misclassification} & VGG, Inception,\\
& & & DenseNet, ResNet,\\
& & & AlexNet, etc. \\\hline
\cite{rakin2019bit} & bit flip & misclassification & AlexNet, ResNet  \\\hline
\cite{rakin2021t} & bit flip & misclassification & AlexNet, MobileNet  \\\hline
\multirow{3}{*}{\cite{breier2021sniff}} & \multirow{3}{*}{bit flip} &\multirow{3}{*}{ model extraction} & VGG, ResNet,\\
& & & DenseNet, Inception,\\
& & & GoogleNet, etc.  \\\hline
\cite{rakin2020tbt} & bit flip & trojan insertion & ResNet, VGG  \\\hline
\cite{breier2022foobar} & instruction skip & backdoor injection & MLP, CNN  \\\hline
\cite{luo2021deepstrike} & timing violations & misclassification & LeNet-5   \\\hline
\cite{chen2021proflip} & bit flip & trojan insertion & ResNet, VGG  \\\hline
\cite{bai2021targeted} & bit flip & misclassification & ResNet, VGG  \\\hline
\end{tabular}
\vspace{3mm}
\caption{\label{tab:fia_dl}Summary of the state-of-the-art fault injection attacks on neural networks. }
\end{center}
\end{table}

\subsection{Protection Techniques Against Fault Attacks}

\subsubsection{Countermeasures for Cryptographic Implementations}
In the area of cryptography, many strategies have been proposed to date to protect the implementations against fault attacks.
The most straightforward way is temporal and spatial duplication: in the case of the temporal one, the algorithm is executed twice and the result is compared; in the case of the spatial duplication, the entire circuit is deployed twice.
While such a strategy can protect against a single fault, it was shown that with a precise double fault injection, this type of protection can be rendered useless~\cite{selmke2016attack}.
Instruction-based redundancy can be used in software, especially to prevent powerful instruction skip/change attacks~\cite{patrick2017lightweight}.
In hardware, active detection circuits can be implemented to monitor sudden environmental changes that could indicate adversarial action~\cite{zussa2014efficiency,he2017fpga}.
Then, there are techniques that are specific to cryptography, such as infective countermeasures~\cite{patranabis2015fault}, fresh re-keying~\cite{mennink2020beyond}, and fault-resistant cipher design~\cite{baksi2021default}.
Finally, as an inspiration for this work, several research works focus on utilizing coding theory to protect against malicious faults, either in software~\cite{breier2017feeding,breier2019evaluating}, or hardware~\cite{guo2013recomputing}.

\subsubsection{Countermeasures for Neural Networks}
When it comes to approaches tailored for neural networks, the area is currently gaining momentum.
Below, we mention the most recent approaches.

In RA-BNN~\cite{rakin2021ra}, the authors suggest using binary neural networks (BNNs) to enhance protection against bit-flip attacks.
The resistance against BFA~\cite{rakin2019bit} is claimed to be improved over $125\times$.
This method needs re-training of the entire model and the model size of the robust BNN is $4\times$ bigger than its unprotected counterpart, however, it is still relatively comparable to a $2\!-\!6$ bit quantized network.

To reduce the performance degradation from using BNNs, it was proposed to use a \textit{piece-wise clustering} (PC)~\cite{he2020defending}.
The trade-off is lower resistance to bit-flip attacks compared to~\cite{rakin2021ra}.
The required number of bit-flips for BFA was increased $2.09\times$ for $8-$bit quantized ResNet-20.

RADAR~\cite{li2021radar} protection derives a $2-$bit checksum for a group of weights in memory before the deployment and then checks the integrity of the weights at runtime.
The claimed detection rate is $96.1\%$ with less than $1\%$ computational overhead and $5.6$ KB of memory overhead for ResNet-18.
However, if the attacker targets the checksum itself, this proposal can be easily broken with one additional fault.

A similar idea was utilized in the HASHTAG~\cite{javaheripi2021hashtag} approach, which first hashes the benign DNN before the deployment to get a snapshot of an untampered network and then compares this hash during the runtime.
Compared to RADAR, the HASHTAG detection rate on ResNet-18 is $100\%$ with $0.01\%$ computational overhead and $1.8$ KB of memory overhead.

Aegis~\cite{wang2023aegis} involves early exit strategies from the network that are randomized so that the attacker does not know, which path is being taken.
More specifically, after each layer, an entire classifier is attached that can finalize the network computation.
The decision on how many original layers are executed, and when the dynamic exit is taken, is random.
By this method, the attack success rate is reduced by $5\!-\!10\times$.
For large models, the overheads are at an acceptable level.
However, for small networks the space overhead is over $3\times$.

NeuroPots~\cite{liu2023neuropots} uses an approach adopted from \textit{HoneyPots} security mechanism that is used to lure the attacker into focusing on something that pretends to be a vulnerable high-reward target, but is in fact planted there by design as a distraction.
With this approach, several ``honey neurons'' and ``honey weights'' are planted in the network based on multiple criteria and then, a detection mechanism is running during runtime to detect tampering with these.
As the number of honey weights is low compared to the network size, the detection overhead is much lower compared to checking the entire network. 
The time overhead stays below $10\%$ and the memory overhead is at most $1.3\%$ for ResNet-20 and it is less for other networks (e.g., VGG-16, AlexNet, ResNet-34).

Output code matching (OCM)~\cite{ozdenizci2022improving} aims at reducing attack stealthiness, and therefore targets the attacks that consider that parameter as their goal --- stealthy TA-LBF~\cite{bai2021targeted}, and stealthy T-BFA~\cite{rakin2021t}.

DNN-Defender~\cite{zhou2023dnn} leverages in-DRAM swapping to protect against Rowhammer-based bit flips.
It downgrades targeted attack performance to the level of a random attack.
It is a hardware-based defense with overheads in terms of latency and area.

%DRAM-Locker~\cite{zhou2023dram}

% \begin{table}[tb]
%     \centering\small
%     \begin{tabular}{|c|c|} \hline
%         Work & Security\\ \hline\hline
%         RA-BNN~\cite{rakin2021ra} & $\times 125$ bit flips \\\hline
%         PC~\cite{he2020defending} & $\times 2.09$ bit flips\\\hline
%         RADAR~\cite{li2021radar} & $96.1\%$ detection rate\\\hline
%         HASHTAG~\cite{javaheripi2021hashtag} & $100\%$ detection rate\\\hline
%         Aegis~\cite{wang2023aegis} & $\times 5-\times 10$ reduced success rate\\\hline
%         NeuroPots~\cite{liuneuropots} & $100\%$ detection rate \\\hline
%         OCM~\cite{ozdenizci2022improving} & \\\hline
%         DNN Defender~\cite{zhou2023dnn} & \\\hline
%         DNN Locker~\cite{zhou2023dram} & \\\hline
%     \end{tabular}
%     \vspace{3mm}
%     \caption{Protection techniques against fault attacks on neural networks. \jb{finalize later}}
%     \label{tab:background_fia_counter}
% \end{table}

\subsubsection{Comparison to existing countermeasures}
RA-BNN~\cite{rakin2021ra} and Aegis~\cite{wang2023aegis} incur a relatively high overhead compared to \method: $3-4\times$ the model size.
The most efficient scenario in our case needs $50\%$ more memory and the least efficient $125\%$.
OCM~\cite{ozdenizci2022improving} only protects against stealthiness and does not consider general-type attacks.
Piece-wise clustering~\cite{he2020defending} aims at reducing the performance degradation and only provides a marginal improvement in terms of the increased number of bit-flips.
RADAR~\cite{li2021radar} requires a secure on-chip SRAM to store certain protection parameters.
Moreover, the premise is the attacker does not flip the $2-$bit checksum, thus placing a relatively strong assumption on the attacker model.
Together with HASHTAG~\cite{javaheripi2021hashtag}, these two protection methods focus on detecting bit-flips instead of increasing the attacker's effort.
The overall direction is therefore orthogonal to \method.
NeuroPots\cite{liu2023neuropots} is a completely different approach to detecting bit-flips by altering the model parameters, which involves an accuracy drop.
DNN-Defender~\cite{zhou2023dnn} is a hardware-based method and can be used on top of the software approaches to add an additional layer of security.

\section{Threat Model and Assumptions}
\label{sec:assumptions}
\method countermeasure proposed in this work protects against bit-flip attacks that target model parameters stored in the memory.
Our proposal is universally applicable to protect against any fault injection technique, such as Rowhammer, laser, electromagnetic pulse, etc.
As the majority of the previous work only considers Rowhammer-based attacks, we aim our discussions in the rest of this paper mainly at that attack vector.
There is no specific assumption on how the data is stored in the memory.

\noindent
\textbf{Hardware threat model.}
We assume a white-box knowledge of the DRAM memory mapping, i.e., the attacker is capable of locating the physical addresses of the target model parameters.
We also assume the attacker can flip any chosen bit in the data memory -- we do not have any assumption on in-memory protection, such as checksums or hashes.
On the other hand, we assume that the decoding routine is not tampered with by the attacker.
We argue this is a reasonable assumption -- this routine is part of the program, and therefore, stored in the program memory.
Program memory is typically a non-volatile flash memory (as opposed to data memory which is a volatile DRAM), and therefore, safe against Rowhammer and most of the other fault injection methods.
This is in contrast to approaches utilizing checksums or hashes that would be stored in the DRAM.

\noindent
\textbf{Software threat model.}
We assume a white-box threat model for bit-flip attacks widely utilized in previous works (e.g., in~\cite{rakin2019bit,rakin2021t,bai2021targeted}).
In this model, the attacker has knowledge of all the model details, such as architecture, model parameter values, quantization width, etc.
At the same time, the attacker can run user-level processes on the target machine, allowing them to mount a Rowhammer attack successfully.

To summarize, we allow the worst-case scenario attacker knowledge and power from the defender's point of view and do not impose any specific assumptions on secure hardware, such as a trusted execution environment or hardware security module.

\section{Methods}
\label{sec:method}

In this section, we first give details on weight quantization and how the numbers are represented in memory (subsection~\ref{sec:quantization}). 
Then in subsection~\ref{sec:bitflip-statistics} we further analyze how the weight values are changed during the bit flip attacks.
Finally, in subsection~\ref{sec:choiceofcode} we provide a theoretical background on binary codes and explain our code choices for the countermeasure.

\subsection{Weight Quantization and Two's Complement Encoding}
\label{sec:quantization}
Following~\cite{rakin2019bit,rakin2021t,bai2021targeted}, we consider a layer-wise $b-$bit (normally $b=2,4,8,16$) uniform post-training quantization scheme for weight quantization.
In particular, a given weight parameter $\omega$ is quantized to a $b-$bit integer $v\in\{0,1\}^b$ such that
\[
\omega=v\Delta,
\]
where $\Delta>0$ is a parameter identical for all weights in a given layer.

Instead of storing the floating point value of the weight $\omega$, the quantized value $v$ is stored together with the parameter $\Delta$.
Since $\omega$ is normally either $32-$bit or $64-$bit and $v$ is $b-$bit, a significant amount of memory can be saved.

Integers $v$ are stored in two's complement encoding.
More specifically, the string $(v_1,v_2,\dots,v_b)\in\{0,1\}^b$ represents the integer
\[
-2^{b-1} v_b+\sum_{i=1}^{b-1}2^{i-1}v_i.
\]

\begin{table}[tb]
    \centering\small
    \begin{tabular}{|c|c||c|c|}\hline
     7 & 0111 & -1 & 1111\\
     6 & 0110 & -2 & 1110\\
     5 & 0101 & -3 & 1101\\
     4 & 0100 & -4 & 1100\\
     3 & 0011 & -5 & 1011\\
     2 & 0010 & -6 & 1010\\
     1 & 0001 & -7 & 1001\\
     0 & 0000 & -8 & 1000\\
     \hline
    \end{tabular}
    \vspace{3mm}
    \caption{Two's complement encoding for 4-bit quantized values.}
    \label{tab:twoscomp}
\end{table}

We can see that the value $b$, called the \textit{quantization bit width}, specifies that each quantized value will be an integer in the range $[-2^{b-1},2^{b-1}-1]$.
For $b=4$, integers from $-8$ to $7$ and their corresponding two's complement encodings are shown in Table~\ref{tab:twoscomp}.

In this paper, we will focus on the cases $b=4$ and $b=8$.

Most of the proposed bit flip attacks aim to flip bits in the quantized values $v$ in the memory before the inference computation, resulting in certain attacker-specified behavior of the network.
Bits that are supposed to be flipped are carefully selected beforehand by comparing the network behavior before and after the bit flips.
Different numbers of bit flips are needed for different changes in the quantized weight values.
For example, according to Table~\ref{tab:twoscomp}, if the attacker would like to change a value from $7$ to $6$, only a single bit flip is needed on the least significant bit.
And to change $-7$ to $-6$, two bit flips are needed.
In Table~\ref{tab:twos}, we list the number of bit flips required to change one integer to another for $4-$bit quantized integers (from $-8$ to $7$).

\begin{table}[tb]
    \centering
    \setlength{\tabcolsep}{4pt}
    %\small
    \begin{tabular}{|c|cccccccccccccccc|}\hline
       & -8 & -7 & -6 & -5 & -4 & -3 & -2 & -1 & 0 & 1 & 2 & 3 & 4 & 5 & 6 & 7 \\\hline
    -8 & - & 1 & 1 & 2 & 1 & 2 & 2 & 3 & 1 & 2 & 2 & 3 & 2 & 3 & 3 & 4 \\\hline
    -7 & 1 & - & 2 & 1 & 2 & 1 & 3 & 2 & 2 & 1 & 3 & 2 & 3 & 2 & 4 & 3 \\\hline
    -6 & 1 & 2 & - & 1 & 2 & 3 & 1 & 2 & 2 & 3 & 1 & 2 & 3 & 4 & 2 & 3 \\\hline
    -5 & 2 & 1 & 1 & - & 3 & 2 & 2 & 1 & 3 & 2 & 2 & \coften1 & 4 & 3 & 3 & 2 \\\hline
    -4 & 1 & 2 & 2 & 3 & - & 1 & 1 & 2 & 2 & 3 & 3 & 4 & \coften1 & 2 & 2 & 3 \\\hline
    -3 & 2 & 1 & 3 & 2 & 1 & - & 2 & 1 & 3 & 2 & 4 & 3 & 2 & \coften1 & 3 & 2 \\\hline
    -2 & 2 & 3 & 1 & 2 & 1 & 2 & - & 1 & 3 & 4 & 2 & 3 & 2 & 3 & \coften1 & 2 \\\hline
    -1 & 3 & 2 & 2 & 1 & 2 & 1 & 1 & - & 4 & 3 & 3 & 2 & 3 & 2 & 2 & \coften1 \\\hline
    0 & \coften1 & 2 & 2 & 3 & 2 & 3 & 3 & 4 & - & 1 & 1 & 2 & \coften1 & 2 & \coften 2 & \coften 3 \\\hline
    1 & 2 & \coften1 & 3 & 2 & 3 & 2 & 4 & 3 & 1 & - & 2 & 1 & 2 & \coften1 & 3 & \coften 2 \\\hline
    2 & 2 & 3 & \coften1 & 2 & 3 & 4 & 2 & 3 & 1 & 2 & - & 1 & 2 & 3 & \coften1 & 2 \\\hline
    3 & 3 & 2 & 2 & \coften1 & 4 & 3 & 3 & 2 & 2 & 1 & 1 & - & 3 & 2 & 2 & \coften1 \\\hline
    4 & 2 & 3 & 3 & 4 & \coften1 & 2 & 2 & 3 & 1 & 2 & 2 & 3 & - & 1 & 1 & 2 \\\hline
    5 & 3 & 2 & 4 & 3 & 2 & \coften1 & 3 & 2 & 2 & 1 & 3 & 2 & 1 & - & 2 & 1 \\\hline
    6 & 3 & 4 & 2 & 3 & 2 & 3 & \coften1 & 2 & 2 & 3 & 1 & 2 & 1 & 2 & - & 1 \\\hline
    7 & 4 & 3 & 3 & 2 & 3 & 2 & 2 & \coften1 & 3 & 2 & 2 & 1 & 2 & 1 & 1 & - \\\hline
    \end{tabular}
    \vspace{3mm}
    \caption{Number of bit flips needed to change between each pair of integers from $-8$ to $7$ when they are represented using two's complement encoding (see Table~\ref{tab:twoscomp} for the encoding list).}
    \label{tab:twos}
\end{table}

\subsection{Bit Flip Statistics}
\label{sec:bitflip-statistics}
Before we discuss how the countermeasure is designed, we will provide a more detailed analysis of how the weights are changed during the bit flip attacks.

\begin{figure}
    \centering
    \setlength{\tabcolsep}{0pt}
    \begin{tabular}{cc}
     \includegraphics[width=0.24\textwidth]{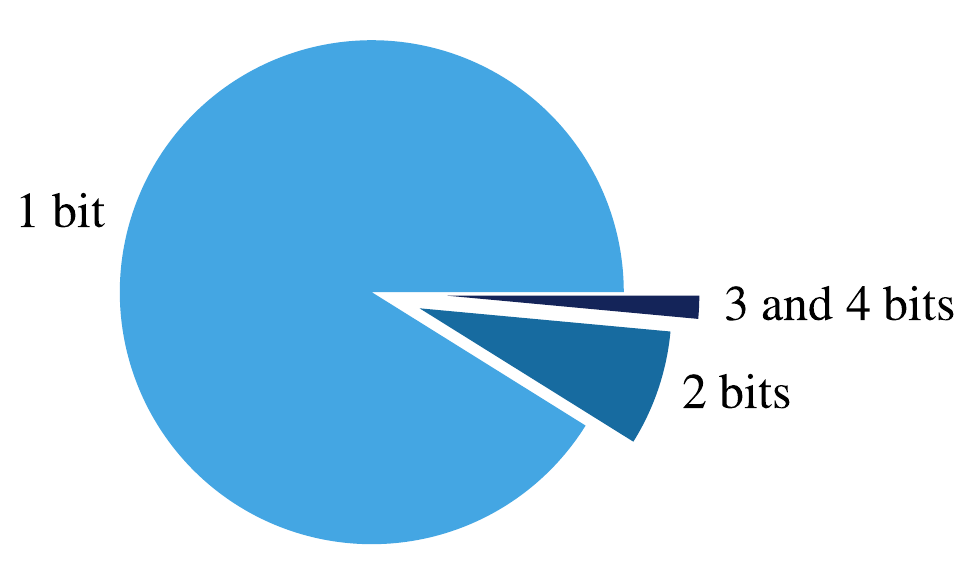} &
    \includegraphics[width=0.24\textwidth]{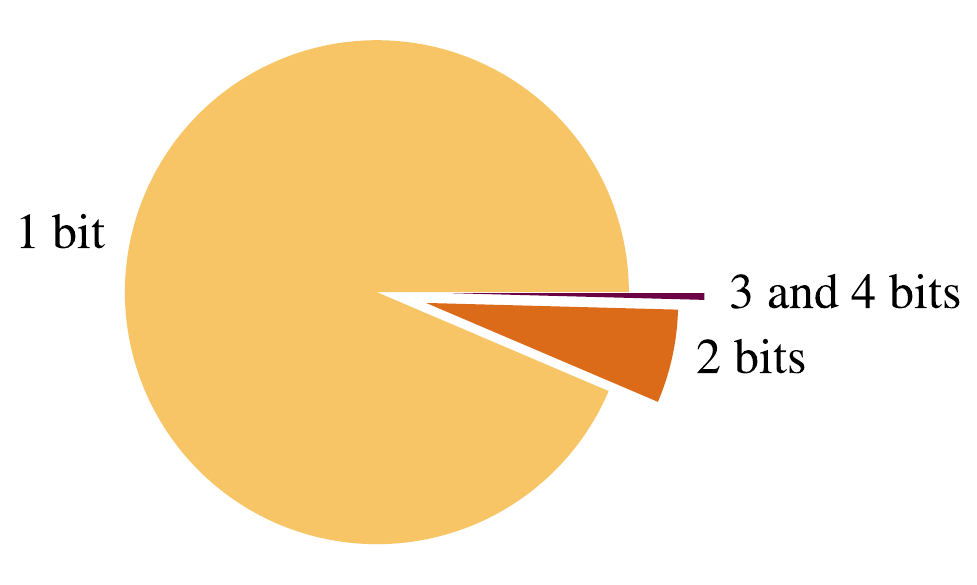}\\
    {\footnotesize 4-bit quantization} & {\footnotesize 8-bit quantization}
    \end{tabular}
    \caption{
    $7,500$ different attacks based on three recent bit flip attacks on quantized neural networks were carried out.
    Those attacks result in $\later{147,106}$ weight value changes for $4-$bit quantized neural networks and $\later{158,310}$ weight value changes for $8-$bit quantized neural networks.
    The figures illustrate the percentage of the number of bits attacked per weight value.
    Note that numbers for 4 bits are too small to show on the chart ($\approx 0.01\%$ for 4-bit quantization and $\approx 0.003\%$ for 8-bit quantization), therefore they were merged with 3 bits.}
    \label{fig:bit_change}
\end{figure}

First, we want to find out how many bits will be flipped when the weight values are changed.
In Figure~\ref{fig:bit_change} we summarize the analysis with three recent bit-flip attacks on $4-$bit and $8-$bit quantized neural networks~\cite{rakin2019bit,rakin2021t,bai2021targeted}.
The detailed experimental settings can be found in Section~\ref{sec:eval}, where attacks were carried out for each attack method and each network structure, in total $\later{3,800}$ different attacks for $4-$bit quantized networks and $\later{3,700}$ attacks for $8-$bit quantized networks.
Those attacks result in $\later{147,106}$ weight value changes for $4-$bit quantized networks and $\later{158,310}$ weight value changes for $8-$bit quantized networks.
We see from Figure~\ref{fig:bit_change} that for $4-$bit quantized networks, most attacks result in a one-bit change in the quantized weight value, some cause two-bit changes, and very few are three or four-bit changes.
Similarly, for $8-$bit quantized networks, most attacks change one or two bits in the quantized weight value, only a few change three or four bits, and no attack changes more than four bits.

\begin{figure}[tb]
    \centering
    \includegraphics[trim={0 3cm 1.5cm 0.7cm},clip,width=0.49\textwidth]{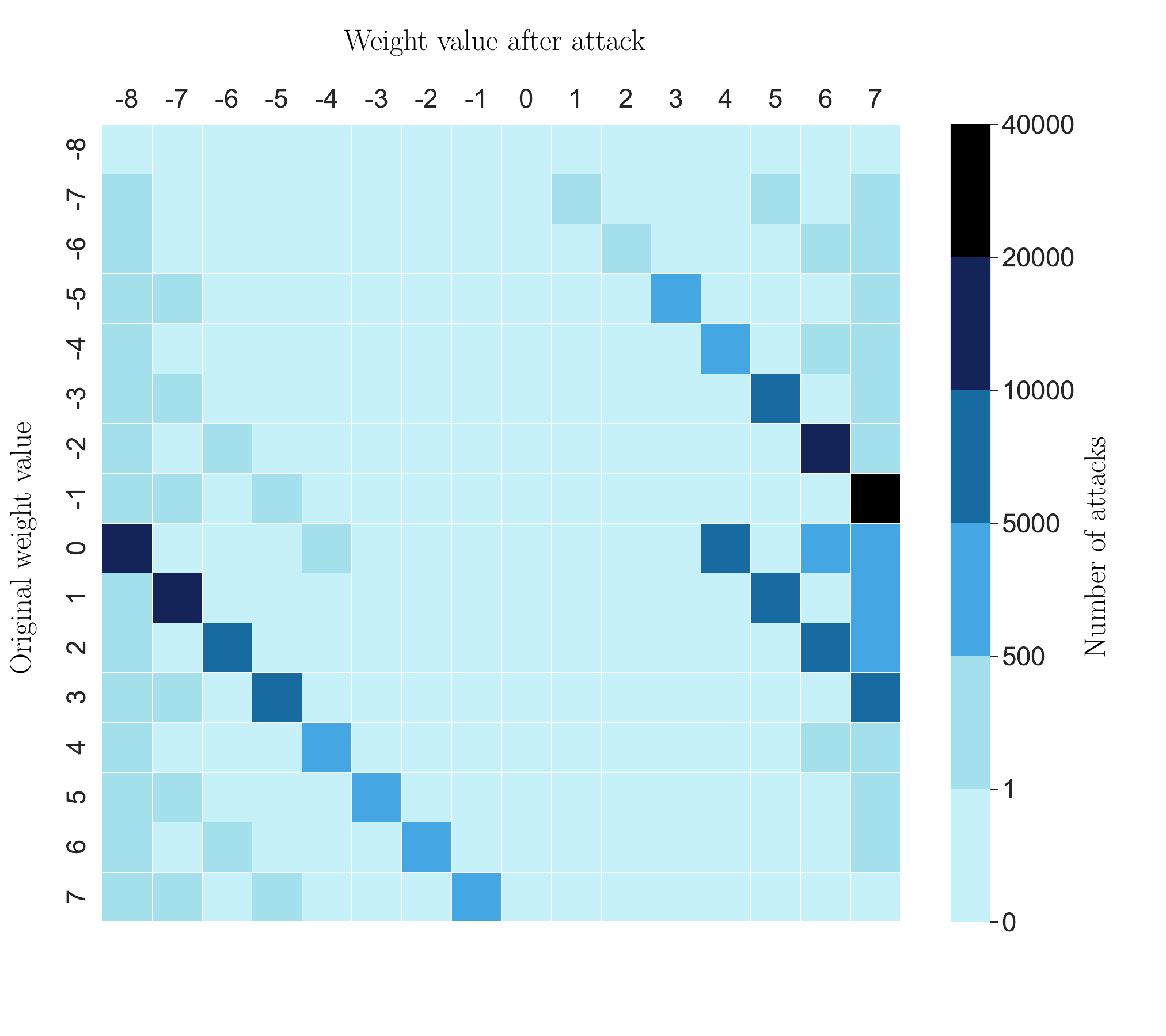}
    \caption{$3,800$ different attacks based on three recent bit flip attacks on $4$-bit quantized neural networks were carried out.
    Those attacks result in $\later{147,106}$ weight value changes.
    The heat map summarizes how often a weight value (rows) is changed to another value (columns) in those attacks.}
    \label{fig:heat_4bit}
\end{figure}

For further analysis, in Figure~\ref{fig:heat_4bit} we summarize how often one integer is changed to another integer for $4-$bit quantized weights.
We can see that the distribution of the attacked bits is concentrated on only certain pairs of integers.
Most pairs are not changed to each other in any of the attacks.
The $20$ most often changed values are also highlighted in blue in Table~\ref{tab:twos}.
Same as in Figure~\ref{fig:heat_4bit}, we consider the rows to correspond to the original values and the columns to correspond to the changed values.
We can see that among those $20$ pairs of quantized weights, only $3$ do not have distance $1$.
Moreover, $13$ of them correspond to changes in the most significant bit (MSB).
In fact, $\later{71.4}\%$ (\later{$105,074$ out of $147,106$}) of the bit flips change the MSB of the weight value.
%MSB: 105074, total: 147106, percentage: 
Similar observations also hold for attacks on $8-$bit quantized networks.
In particular, $\later{80.4}\%$ (\later{$127,204$ out of $158,310$}) of the bit flips change the MSB of the quantized weight values.

\subsection{The Choice of Binary Codes}
\label{sec:choiceofcode}

In this subsection, we provide basic notions regarding binary codes \cite{macwilliams+sloane, ling2004coding} and discuss our choice of binary codes for the proposed countermeasure implementation.

In what follows $n$ denotes a positive integer, $\FF_2=\{0,1\}$ the binary field, and $\FF_2^n$ the set of all binary strings of length $n$.
It is understood that addition in $\FF_2$ is performed modulo 2, i.e., $0+0 = 1+1 = 0$ and $0+1 = 1+0 = 1$.
Equivalently, addition in $\FF_2$ is the familiar XOR operation.
The sum of two strings $\boldsymbol{u}=(u_1, \ldots, u_n), \boldsymbol{v}=(v_1, \ldots, v_n) \in \FF_2^n$ is another string from $\FF_2^n$ obtained as their coordinate-wise modulo $2$ sum, namely $\boldsymbol{u}+\boldsymbol{v}=(u_1+v_1, \ldots, u_n+v_n)$.
For example, $111 + 110 = 001$.

%A binary code is a set of binary strings, formally~\cite[page~5]{ling2004coding}:
\begin{definition}[{\cite[Section 2.1]{ling2004coding}}]
A nonempty set $C\subseteq\FF_2^n$ is called a \textit{binary code} of \textit{length} $n$.
Elements of $C$ are called \textit{codewords}.
%The number of codewords in $C$ is called the \textit{size} of $C$.
A binary code of length $n$ and cardinality $M$ is called an $(n,M)-$binary code.
\end{definition}
For example, $\{00,11\}\subseteq\FF^2_2$ is a
%binary code of length $2$ and size $2$.
$(2,2)-$binary code.

As mentioned in subsection~\ref{sec:quantization}, in our case study we will look at $4-$bit and $8-$bit quantizations.
The weight values are between $-8$ and $7$ (resp. $-128$ and $127$).
In total, there are $16$ (resp. $256$) possible values for quantization bit width $b=4$ (resp. $b=8$).
To be able to find a codeword for each of the weight values, we need a binary code of cardinality $16$ for $b=4$ and $256$ for $b=8$.

When the intermediate values of a computation are encoded, an attack within the bit flip fault model includes flipping one or more bits of a codeword.
In general, the resulting string may or may not be a codeword.
Codes having the property that $t$ or fewer bit flips in an arbitrary codeword never result in another codeword are said to be $t-$error \textit{detecting}.
E.g., the code $C=\{00,11\}$ from the above example is $1-$error detecting because one bit flip in a codeword will produce either $01$ or $10$, neither of which is in $C$.

Error-detecting codes are needed in our context because we aim to increase the number of bit flips that are necessary for the attacks.
For example, if a code is $1-$error detecting, then changing one codeword into another will require more than one bit flip.
As we discussed in subsection~\ref{sec:bitflip-statistics}, most of the weight value changes will have $1$ or $2$ bit flips. Hence, we want codes that can detect at least $2$ errors.
We next introduce the notion of Hamming distance between binary strings and the minimum distance of a code, through which the error detection capability of a code is most naturally expressed.

The Hamming distance between two strings $\boldsymbol{u}=(u_1, \ldots, u_n), \boldsymbol{v}=(v_1, \ldots, v_n) \in \FF_2^n$, denoted $\dis{\boldsymbol{u},\boldsymbol{v}}$, is the number of coordinates at which they differ, or formally
\begin{subequations}
\begin{equation}
\dis{\boldsymbol{u},\boldsymbol{v}}=\sum_{i=1}^{n}\dis{u_i,v_i},
\end{equation}
where
\begin{equation}
\dis{u_i,v_i}=\begin{cases}
    1 & \text{ if } u_i\neq v_i\\
    0 & \text{ if } u_i=v_i
    \end{cases}.
\end{equation}
\end{subequations}
For example,
\[
\dis{01,10}=2,\ \dis{111,110}=1,\ \dis{010101,101010}=6.
\]
The Hamming distance between $\boldsymbol{u}$ and the string of all zeros $\boldsymbol{0}=(0,\ldots,0)$, namely $\dis{\boldsymbol{u}, \boldsymbol{0}}$, equals the number of $1$'s in $\boldsymbol{u}$ and is called the Hamming weight of $\boldsymbol{u}$.
It is easy to see that the Hamming distance between $\boldsymbol{u}$ and $\boldsymbol{v}$ equals the Hamming weight of $\boldsymbol{u} + \boldsymbol{v}$.

The \textit{minimum distance}~\cite[Section 2.5]{ling2004coding} of a binary code $C$, denoted $\dis{C}$, is the smallest Hamming distance between any two distinct codewords, i.e.,
\begin{equation}
    \dis{C}=\min\{\dis{\boldsymbol{c}_1,\boldsymbol{c}_2} \;|\; \boldsymbol{c}_1,\boldsymbol{c}_2\in C, \boldsymbol{c}_1\neq\boldsymbol{c}_2\}.
\end{equation}
An $(n,M)-$binary code of minimum distance $d$ is called an $(n,M,d)-$\allowbreak binary code.
For example, $\{00,11\}\subseteq\FF_2^2$ is a $(2,2,2)-$binary code, and $\{0000,1111,1110\}\subseteq\FF_2^4$ is a $(4,3,1)-$binary code.
A binary code of minimum distance $d$ is $(d-1)-$error detecting \cite[Theorem 2.5.6]{ling2004coding}.

%Now, suppose $C$ is a binary code that has a minimum distance $1$.
%By definition, there exists $\boldsymbol{c}_1,\boldsymbol{c}_2\in C$ such that $\dis{\boldsymbol{c}_1,\boldsymbol{c}_2}=1$.
%If $\boldsymbol{c}_1$ is used during the computation and a single-bit flip happens, we might get $\boldsymbol{c}_2$ and we will not be able to tell if there was an error during the computation.

Based on the above-mentioned requirements, what we need are $(n,M,d)-$binary codes with $M \in \{16, 256\} $ and $d\geq3$.
Further, for $4-$bit quantization we set $n\leq 9$, and for $8-$bit quantization we set $n\leq 14$.
Naturally, it is possible to consider longer codes, but with current neural network architectures that already require large amounts of memory, the overheads would be impractical.
Furthermore, for each of the considered $(n,M)$ pairs, we would like to find binary codes that achieve the maximum possible value of $d$.
These maximum values can easily be inferred from elementary bounds in coding theory~\cite[Section 5]{ling2004coding}, and are listed in Table~\ref{tab:codes}.

\begin{table}[tb]
    \centering\small
    \begin{tabular}{|c|c|c|c|c|}\hline
   $M$ &  $n$   &  $d^*$ & Chosen code  & Notation\\\hline\hline
   &  $7$   & $3$ & Hamming $(7,2^4)$ code & $C_{(7,3)}$\\\cline{2-5}
  $16$ &  $8$   & $4$ & extended Hamming $(8,2^4)$ code & $C_{(8,4)}$\\\cline{2-5}
   &  $9$   & $4$ & \makecell{extended Hamming $(16,2^{11})$\\code shortened at $7$ locations} & $C_{(9,4)}$\\\hline
   &  $12$   & $3$ & \makecell{Hamming $(15,2^{11})$ code\\shortened at $3$ locations} & $C_{(12,3)}$\\\cline{2-5}
  $256$ &  $13$   & $4$ & \makecell{extended Hamming $(16,2^{11})$\\code shortened at $3$ locations} & $C_{(13,4)}$\\\cline{2-5}
   &  $14$   & $4$ & \makecell{subcode of extended Hamming\\ $(16,2^{11})$ code shortened at $2$ loc.} & $C_{(14,4)}$\\\hline
    \end{tabular}
    \vspace{3mm}
    \caption{For each code size $M$ and length $n$, the maximum possible code distance $d^*$ is shown in the third column.
    In the fourth column, we list our chosen binary codes that achieve the given parameters.
    The last column lists the notation used in the rest of this paper for each of these codes.}
    \label{tab:codes}
\end{table}

The codes we chose in this study are based on the following two well-known families \cite[Section 5.3.1]{ling2004coding}:
\begin{itemize}
\item Hamming codes -- binary codes of length $2^r-1$, cardinality $2^{2^r-r-1}$, and minimum distance $3$, and
\vspace{2mm}
\item Extended Hamming codes -- binary codes of length $2^r$, cardinality $2^{2^r-r-1}$, and minimum distance $4$, obtained from Hamming codes by adding an overall parity check bit (i.e., the XOR of the first $2^r-1$ bits) to all codewords.%
\end{itemize}
Here $r$ is a parameter that can take on any integral value larger than $1$.
These codes are known to be optimal for their respective parameters, i.e., there exists no larger code of length $2^r-1$ (resp. $2^r$) and minimum distance $3$ (resp. $4$).
In fact, Hamming codes are \emph{perfect} in the sense that every binary string of length $2^r-1$ is either a codeword or is at distance $1$ from exactly one codeword.
These and other mathematical properties of Hamming and extended Hamming codes make them particularly convenient for analysis.

For example, the case $r=3$ corresponds to the $(7, 16, 3)$ Hamming code and the $(8, 16, 4)$ extended Hamming code, which are denoted in this paper by $C_{(7,3)}$ and $C_{(8,4)}$, respectively.
We list here explicitly the codewords of the former:
%\begin{equation}
\begin{alignat}{4}
\nonumber
C_{(7,3)} = \big\{ &0000000 ,\;\; &&1001011 ,\;\; &&0010111 ,\;\; &&1011100 , \\
\nonumber
&1100101 , &&0101110 , &&1110010 , &&0111001 , \\
\nonumber
&1111111 , &&0110100 , &&1101000 , &&0100011 , \\
\label{eq:C73}
&0011010 , &&1010001 , &&0001101 , &&1000110 \big\} .
\end{alignat}
%\end{equation}

In order to obtain the remaining codes with parameters that do not correspond to either of the above two families, we employ the operation of shortening.
Shortening a code $C$ at the $i$'th location produces a code $C_i$ defined as follows: we keep only the codewords of $C$ that have $0$ as the $i$'th coordinate and then delete the $i$'th coordinate from all of these codewords \cite[page 126]{ling2004coding}. It is easy to see that this operation cannot decrease the minimum distance, i.e., $\dis{C_i} \geq \dis{C}$.
For example, for the code $C = \{0000,1001,1110\} \subseteq \FF_2^4$, we would have $C_1 = \{000\}$, $C_2 = C_3 = \{000, 101\}$, $C_4=\{000, 111\}$.

Due to the structure of the Hamming codes, one can show that, e.g., shortening the $(16, 2^{11}, 4)$ extended Hamming code at the last $7$ locations produces a $(9, 2^4, 4)$-binary code, which we denote herein by $C_{(9,4)}$.
Likewise, shortening the $(15, 2^{11}, 3)$ Hamming code (resp. $(16, 2^{11}, 4)$ extended Hamming code) at the last $3$ locations produces a $(12, 2^8, 3)-$binary code (resp. $(13, 2^8, 4)-$binary code), denoted by $C_{(12,3)}$ (resp. $C_{(13,4)}$).
Finally, shortening the $(16, 2^{11}, 4)$ extended Hamming code at the last $2$ locations produces a $(14, 2^9, 4)-$binary code.
Our code $C_{(14,4)}$ is a randomly chosen linear subcode (see Definition \ref{def:linearcode} below) of size $2^8 = 256$ of the latter code.

Having a set of $M \in \{16, 256\}$ codewords that we intend to use, it is left to specify which integer value is encoded to which codeword.
It is easy to see that if the distance between two codewords is big, more bit flip attacks will be needed to transform one codeword into another.
Thus, to decide how to encode the integer values, we need to find out which pairs of integers are more likely to be changed to each other during the attacks and hence should have corresponding codewords with larger distances.
This analysis was already done in subsection~\ref{sec:bitflip-statistics} where we found that, in most cases, the attacks only change one or two bits in the weight values.
Moreover, among those $1-$bit flips, the majority focus on changing the MSB of the quantized values.
Consequently, the integers that only differ in the MSB should be mapped to codewords with the largest possible distances.

To describe in more precise terms the encoding we chose, another notion from coding theory will be useful.
Recall that $\FF_2^n$ can be regarded as an $n$-dimensional vector space over $\FF_2$ \cite[Section 4.1]{ling2004coding}.

\begin{definition}[{\cite[Section 4.2]{ling2004coding}}]
\label{def:linearcode}
A binary code $C \subseteq \FF_2^n$ is called linear if it is a vector subspace of the space $\FF_2^n$. Equivalently, $C$ is linear if the sum of any two (not necessarily different) codewords $\boldsymbol{c}_1, \boldsymbol{c}_2 \in C$ is also a codeword.
\end{definition}

For example, Hamming and extended Hamming codes are linear. Furthermore, shortening a linear code produces a linear code and, hence, all codes listed in Table~\ref{tab:codes} are linear.

Let $\boldsymbol{e}^{(i)} \in \FF_2^b$ denote the string/vector whose $i$'th coordinate is $1$ and the remaining coordinates are $0$.
The set $\{\boldsymbol{e}^{(1)}, \ldots, \boldsymbol{e}^{(b)}\}$ is the standard basis of the vector space $\FF_2^b$.
In particular, any other element of $\FF_2^b$ can be represented as a linear combination of the $\boldsymbol{e}^{(i)}$'s with coefficients from $\FF_2$.
Namely, for $\boldsymbol{v} = (v_1, \ldots, v_b) \in \FF_2^b$, we have
\begin{equation}
\label{eq:lincomb}
\boldsymbol{v} = \sum_{i=1}^b v_i \boldsymbol{e}^{(i)} .
\end{equation}

If a linear code $C$ is a $b$-dimensional subspace of $\FF_2^n$, its cardinality is $2^b$.
The encoding operation for such a code is a linear mapping $f_C \colon \FF_2^b \to \FF_2^n$ that specifies which codeword of $C$ is assigned to which ``information string'' $\boldsymbol{v} \in \FF_2^b$. The mapping being linear means that
\begin{equation}
\label{eq:linmap}
f_C(\boldsymbol{u}+ \boldsymbol{v}) = f_C(\boldsymbol{u}) + f_C(\boldsymbol{v}) .
\end{equation}
Such a mapping is fully specified by the codewords $f_C(\boldsymbol{e}^{(i)})$ assigned to the strings $\boldsymbol{e}^{(i)}$, $i = 1, \ldots, b$.
This is because the codeword assigned to any other string $\boldsymbol{v}=(v_1, \ldots, v_b)$ can then be determined uniquely from the relation \eqref{eq:lincomb} and the linearity property \eqref{eq:linmap}, namely
\begin{equation}
\label{eq:fC}
f_C(\boldsymbol{v}) = \sum_{i=1}^b v_i f_C(\boldsymbol{e}^{(i)}) .
\end{equation}

Further note that, if two strings $\boldsymbol{u}, \boldsymbol{v} \in \FF_2^b$ differ only at the $i$'th coordinate, implying that their Hamming distance is $1$, then it holds that $\boldsymbol{u} = \boldsymbol{v} + \boldsymbol{e}^{(i)}$.
In particular, if they differ only at the MSB, i.e., at the first coordinate, we have $\boldsymbol{u} = \boldsymbol{v} + \boldsymbol{e}^{(1)}$.
Then, again by the linearity property of $f_C$, we have $f_C(\boldsymbol{u}) = f_C(\boldsymbol{v}) + f_C(\boldsymbol{e}^{(i)})$.
Consequently, the Hamming distance between $f_C(\boldsymbol{u})$ and $f_C(\boldsymbol{v})$ is equal to the Hamming weight of $f_C(\boldsymbol{u}) + f_C(\boldsymbol{v}) = f_C(\boldsymbol{e}^{(i)})$.
This simple observation is important for our chosen encoding operation, so we emphasize it again:
\begin{equation}
\label{eq:distance}
\boldsymbol{u} = \boldsymbol{v} + \boldsymbol{e}^{(i)} \quad \Longrightarrow \quad \dis{f_C(\boldsymbol{u}), f_C(\boldsymbol{v})} = \dis{f_C(\boldsymbol{e}^{(i)}), \boldsymbol{0}} .
\end{equation}

Recall from the earlier discussion that we would like to make the Hamming distance between $f_C(\boldsymbol{u})$ and $f_C(\boldsymbol{v})$ large whenever $\dis{\boldsymbol{u}, \boldsymbol{v}} = 1$.
Based on what was said in the previous two paragraphs, this can be achieved by choosing the codewords $f_C(\boldsymbol{e}^{(i)})$ to be linearly independent and to have large Hamming weights.
In particular, we shall choose for $f_C(\boldsymbol{e}^{(1)})$ a string of the largest possible weight, which by \eqref{eq:distance} implies that whenever two original $b$-tuples differ at the MSB, the corresponding codewords will be at the largest possible distance.

Let us illustrate the above-described codeword assignment for the code $C_{(7,3)}$. As can be seen from \eqref{eq:C73}, the codewords have weights $0, 3, 4, 7$. Since the maximum-weight codeword is $1111111$, we set
\[
f_{C_{(7,3)}}(1000) = 1111111 \,.
\]
We further set
\begin{align*}
f_{C_{(7,3)}}(0100) &= 1100101 \,,  \\
f_{C_{(7,3)}}(0010) &= 0010111 \,,  \\
f_{C_{(7,3)}}(0001) &= 1001011 \,.
\end{align*}
The remaining $12$ codeword assignments are then determined from the relation \eqref{eq:fC}.
For example,
\begin{align*}
f_{C_{(7,3)}}(1101) &= f_{C_{(7,3)}}(1000) + f_{C_{(7,3)}}(0100) + f_{C_{(7,3)}}(0001)  \\ &= 1111111 + 1100101 + 1001011  \\
&=1010001 \,.
\end{align*}
The entire encoding mapping for the code $C_{(7,3)}$, as well as those for the codes $C_{(8,4)}$ and $C_{(9,4)}$, are given in Table~\ref{tab:extham_codewords} (the codewords are represented in hexadecimal).

{\small
\noindent
\begin{table}
\setlength{\tabcolsep}{3pt}
    \centering
    \begin{tabular}{cccccccccccccccc}
        \multicolumn{16}{c}{7-bit Hamming code $C_{(7,3)}$}\\\hline
         \texttt{7F} & \texttt{34} & \texttt{68} & \texttt{23} & \texttt{1A} & \texttt{51} & \texttt{0D} & \texttt{46} & \texttt{00} & \texttt{4B} & \texttt{17} & \texttt{5C} & \texttt{65} & \texttt{2E} & \texttt{72} & \texttt{39}\\\hline\\\\
        \multicolumn{16}{c}{8-bit extended Hamming code $C_{(8,4)}$}\\\hline
         \texttt{FF} & \texttt{B4} & \texttt{E8} & \texttt{A3} & \texttt{9A} & \texttt{D1} & \texttt{8D} & \texttt{C6} & \texttt{00} & \texttt{4B} & \texttt{17} & \texttt{5C} & \texttt{65} & \texttt{2E} & \texttt{72} & \texttt{39}\\\hline\\\\
    \end{tabular}
    \begin{tabular}{cccccccc}
         \multicolumn{8}{c}{9-bit linear code $C_{(9,4)}$}\\\hline
         \texttt{1EF} & \texttt{1F0} & \texttt{193} & \texttt{18C} & \texttt{155} & \texttt{14A} & \texttt{129} & \texttt{136} \\ \texttt{000} & \texttt{01F} & \texttt{07C} & \texttt{063} & \texttt{0BA} & \texttt{0A5} & \texttt{0C6} & \texttt{0D9}\\\hline
    \end{tabular}
    \vspace{4mm}
    \caption{Encoding lists for the chosen codes of length $7, 8, 9$. Codewords are represented in hexadecimal. The first codeword is used to encode $-8$, the second to encode $-7$, and so on.
    For example, using the 7-bit Hamming code, $-5$ is encoded as \texttt{23}.}
    \label{tab:extham_codewords}
\end{table}}

Now, similarly to Table~\ref{tab:twos}, we can compute the number of bit flips required to change one value to another when the integers are encoded in this way.
The details are shown in Tables~\ref{tab:hamming}, ~\ref{tab:exthamming}, and~\ref{tab:linear94} for $C_{(7,3)}$, $C_{(8,4)}$, and $C_{(9,4)}$, respectively.
The same $20$ most often attacked values are highlighted in blue, assuming the rows correspond to the original values and the columns correspond to the changed values.
Compared to Table~\ref{tab:twos}, we can see in those $20$ attacks, the original single-bit flips are replaced by $4,5,7,8$ bit flips, potentially increasing the attack complexities. 
$2-$bit flip attack now require $4-$bit flips and the $3-$bit flip will need either $4-$ or $5-$bit flips.
Furthermore, for integers that differ in the MSB, the distances between the corresponding codewords are either $7$ or $8$.
As mentioned in subsection~\ref{sec:bitflip-statistics}, $\later{71.4}\%$ percent bit flips are on the MSB.
Thus, it is reasonable to assume there will be around $6$ times increase in the number of bit flips with our countermeasure.

\begin{table}[tb]
    \centering
    \setlength{\tabcolsep}{4pt}
    \begin{tabular}{|c|cccccccccccccccc|}\hline
      & -8 & -7 & -6 & -5 & -4 & -3 & -2 & -1 & 0 & 1 & 2 & 3 & 4 & 5 & 6 & 7 \\\hline
-8 & - & 4 & 4 & 4 & 4 & 4 & 4 & 4 & 7 & 3 & 3 & 3 & 3 & 3 & 3 & 3 \\\hline
-7 & 4 & - & 4 & 4 & 4 & 4 & 4 & 4 & 3 & 7 & 3 & 3 & 3 & 3 & 3 & 3 \\\hline
-6 & 4 & 4 & - & 4 & 4 & 4 & 4 & 4 & 3 & 3 & 7 & 3 & 3 & 3 & 3 & 3 \\\hline
-5 & 4 & 4 & 4 & - & 4 & 4 & 4 & 4 & 3 & 3 & 3 & \coften7 & 3 & 3 & 3 & 3 \\\hline
-4 & 4 & 4 & 4 & 4 & - & 4 & 4 & 4 & 3 & 3 & 3 & 3 & \coften7 & 3 & 3 & 3 \\\hline
-3 & 4 & 4 & 4 & 4 & 4 & - & 4 & 4 & 3 & 3 & 3 & 3 & 3 & \coften7 & 3 & 3 \\\hline
-2 & 4 & 4 & 4 & 4 & 4 & 4 & - & 4 & 3 & 3 & 3 & 3 & 3 & 3 & \coften7 & 3 \\\hline
-1 & 4 & 4 & 4 & 4 & 4 & 4 & 4 & - & 3 & 3 & 3 & 3 & 3 & 3 & 3 & \coften7 \\\hline
0 & \coften 7 & 3 & 3 & 3 & 3 & 3 & 3 & 3 & - & 4 & 4 & 4 & \coften 4 & 4 & \coften4 & \coften4 \\\hline
1 & 3 & \coften7 & 3 & 3 & 3 & 3 & 3 & 3 & 4 & - & 4 & 4 & 4 & \coften 4 & 4 & \coften4 \\\hline
2 & 3 & 3 & \coften7 & 3 & 3 & 3 & 3 & 3 & 4 & 4 & - & 4 & 4 & 4 & \coften4 & 4 \\\hline
3 & 3 & 3 & 3 & \coften7 & 3 & 3 & 3 & 3 & 4 & 4 & 4 & - & 4 & 4 & 4 & \coften4 \\\hline
4 & 3 & 3 & 3 & 3 & \coften7 & 3 & 3 & 3 & 4 & 4 & 4 & 4 & - & 4 & 4 & 4 \\\hline
5 & 3 & 3 & 3 & 3 & 3 & \coften7 & 3 & 3 & 4 & 4 & 4 & 4 & 4 & - & 4 & 4 \\\hline
6 & 3 & 3 & 3 & 3 & 3 & 3 & \coften7 & 3 & 4 & 4 & 4 & 4 & 4 & 4 & - & 4 \\\hline
7 & 3 & 3 & 3 & 3 & 3 & 3 & 3 & \coften7 & 4 & 4 & 4 & 4 & 4 & 4 & 4 & - \\\hline
    \end{tabular}
    \vspace{3mm}
    \caption{Number of bit flips needed to change between each pair of integers from $-8$ to $7$ when they are represented using the 7-bit Hamming code $C_{(7,3)}$ (see Table~\ref{tab:extham_codewords} for the encoding list).}
    \label{tab:hamming}
\end{table}

\begin{table}[tb]
    \centering
    \setlength{\tabcolsep}{4pt}
    \begin{tabular}{|c|cccccccccccccccc|}\hline
      & -8 & -7 & -6 & -5 & -4 & -3 & -2 & -1 & 0 & 1 & 2 & 3 & 4 & 5 & 6 & 7 \\\hline
-8 & - & 4 & 4 & 4 & 4 & 4 & 4 & 4 & 8 & 4 & 4 & 4 & 4 & 4 & 4 & 4 \\\hline
-7 & 4 & - & 4 & 4 & 4 & 4 & 4 & 4 & 4 & 8 & 4 & 4 & 4 & 4 & 4 & 4 \\\hline
-6 & 4 & 4 & - & 4 & 4 & 4 & 4 & 4 & 4 & 4 & 8 & 4 & 4 & 4 & 4 & 4 \\\hline
-5 & 4 & 4 & 4 & - & 4 & 4 & 4 & 4 & 4 & 4 & 4 & \coften8 & 4 & 4 & 4 & 4 \\\hline
-4 & 4 & 4 & 4 & 4 & - & 4 & 4 & 4 & 4 & 4 & 4 & 4 & \coften8 & 4 & 4 & 4 \\\hline
-3 & 4 & 4 & 4 & 4 & 4 & - & 4 & 4 & 4 & 4 & 4 & 4 & 4 & \coften8 & 4 & 4 \\\hline
-2 & 4 & 4 & 4 & 4 & 4 & 4 & - & 4 & 4 & 4 & 4 & 4 & 4 & 4 & \coften8 & 4 \\\hline
-1 & 4 & 4 & 4 & 4 & 4 & 4 & 4 & - & 4 & 4 & 4 & 4 & 4 & 4 & 4 & \coften8 \\\hline
0 & \coften8 & 4 & 4 & 4 & 4 & 4 & 4 & 4 & - & 4 & 4 & 4 & \coften4 & 4 & \coften4 & \coften4 \\\hline
1 & 4 & \coften8 & 4 & 4 & 4 & 4 & 4 & 4 & 4 & - & 4 & 4 & 4 & \coften4 & 4 & \coften4 \\\hline
2 & 4 & 4 & \coften8 & 4 & 4 & 4 & 4 & 4 & 4 & 4 & - & 4 & 4 & 4 & \coften4 & 4 \\\hline
3 & 4 & 4 & 4 & \coften8 & 4 & 4 & 4 & 4 & 4 & 4 & 4 & - & 4 & 4 & 4 & \coften4 \\\hline
4 & 4 & 4 & 4 & 4 & \coften8 & 4 & 4 & 4 & 4 & 4 & 4 & 4 & - & 4 & 4 & 4 \\\hline
5 & 4 & 4 & 4 & 4 & 4 & \coften8 & 4 & 4 & 4 & 4 & 4 & 4 & 4 & - & 4 & 4 \\\hline
6 & 4 & 4 & 4 & 4 & 4 & 4 & \coften8 & 4 & 4 & 4 & 4 & 4 & 4 & 4 & - & 4 \\\hline
7 & 4 & 4 & 4 & 4 & 4 & 4 & 4 & \coften8 & 4 & 4 & 4 & 4 & 4 & 4 & 4 & - \\\hline
    \end{tabular}
    \vspace{3mm}
    \caption{Number of bit flips needed to change between each pair of integers from $-8$ to $7$ when they are represented using the 8-bit extended Hamming code $C_{(8,4)}$ (see Table~\ref{tab:extham_codewords} for the encoding list).}
    \label{tab:exthamming}
\end{table}

\begin{table}[tb]
    \centering
    \setlength{\tabcolsep}{4pt}
    \begin{tabular}{|c|cccccccccccccccc|}\hline
       & -8 & -7 & -6 & -5 & -4 & -3 & -2 & -1 & 0 & 1 & 2 & 3 & 4 & 5 & 6 & 7 \\\hline
-8 & - & 5 & 5 & 4 & 5 & 4 & 4 & 5 & 8 & 5 & 5 & 4 & 5 & 4 & 4 & 5 \\\hline
-7 & 5 & - & 4 & 5 & 4 & 5 & 5 & 4 & 5 & 8 & 4 & 5 & 4 & 5 & 5 & 4 \\\hline
-6 & 5 & 4 & - & 5 & 4 & 5 & 5 & 4 & 5 & 4 & 8 & 5 & 4 & 5 & 5 & 4 \\\hline
-5 & 4 & 5 & 5 & - & 5 & 4 & 4 & 5 & 4 & 5 & 5 & \coften8 & 5 & 4 & 4 & 5 \\\hline
-4 & 5 & 4 & 4 & 5 & - & 5 & 5 & 4 & 5 & 4 & 4 & 5 & \coften8 & 5 & 5 & 4 \\\hline
-3 & 4 & 5 & 5 & 4 & 5 & - & 4 & 5 & 4 & 5 & 5 & 4 & 5 & \coften8 & 4 & 5 \\\hline
-2 & 4 & 5 & 5 & 4 & 5 & 4 & - & 5 & 4 & 5 & 5 & 4 & 5 & 4 & \coften8 & 5 \\\hline
-1 & 5 & 4 & 4 & 5 & 4 & 5 & 5 & - & 5 & 4 & 4 & 5 & 4 & 5 & 5 & \coften8 \\\hline
0 & \coften8 & 5 & 5 & 4 & 5 & 4 & 4 & 5 & - & 5 & 5 & 4 & \coften5 & 4 & \coften4 & \coften5 \\\hline
1 & 5 & \coften8 & 4 & 5 & 4 & 5 & 5 & 4 & 5 & - & 4 & 5 & 4 & \coften5 & 5 & \coften4 \\\hline
2 & 5 & 4 & \coften8 & 5 & 4 & 5 & 5 & 4 & 5 & 4 & - & 5 & 4 & 5 & \coften5 & 4 \\\hline
3 & 4 & 5 & 5 & \coften8 & 5 & 4 & 4 & 5 & 4 & 5 & 5 & - & 5 & 4 & 4 & \coften5 \\\hline
4 & 5 & 4 & 4 & 5 & \coften8 & 5 & 5 & 4 & 5 & 4 & 4 & 5 & - & 5 & 5 & 4 \\\hline
5 & 4 & 5 & 5 & 4 & 5 & \coften8 & 4 & 5 & 4 & 5 & 5 & 4 & 5 & - & 4 & 5 \\\hline
6 & 4 & 5 & 5 & 4 & 5 & 4 & \coften8 & 5 & 4 & 5 & 5 & 4 & 5 & 4 & - & 5 \\\hline
7 & 5 & 4 & 4 & 5 & 4 & 5 & 5 & \coften8 & 5 & 4 & 4 & 5 & 4 & 5 & 5 & - \\\hline
    \end{tabular}
    \vspace{3mm}
    \caption{Number of bit flips needed to change between each pair of integers from $-8$ to $7$ when they are represented using the 9-bit code $C_{(9,4)}$ (see Table~\ref{tab:extham_codewords} for the encoding list).}
    \label{tab:linear94}
\end{table}

Similar observations hold for attacks on $8-$bit quantized neural networks.
In that case, we also aim to encode integers that differ in one bit location with codewords having large Hamming distances, by using a procedure analogous to the one described above.
Consequently, we expect to see around $11$ times increase in the number of bit flips for the attacks.

\section{Evaluation}
\label{sec:eval}
To evaluate the effectiveness of our proposed countermeasure, we evaluated three recently proposed bit-flip fault attacks~\cite{rakin2019bit,rakin2021t,bai2021targeted} on several publicly available neural networks.

In subsection~\ref{sec:attacksetting} we will describe the attack methodologies and our choice of datasets as well as network architectures.
Subsection~\ref{sec:evalsettings} will detail the evaluation settings.
The results for unprotected and protected implementations will be presented in subsection~\ref{sec:results}.

\subsection{Attack Settings, Datasets, and Networks}
\label{sec:attacksetting}
First, let us briefly explain the three attack methodologies used for the evaluation in this paper:

\noindent
\textbf{Bit-Flip Attack (BFA)}~\cite{rakin2019bit}: This attack utilizes a method called progressive bit search which combines gradient ranking with progressive search to identify the most vulnerable bits. There is an in-layer and cross-layer bit search that works in an iterative way -- in each iteration, the most vulnerable bit is flipped. 
The goal of the attack is to degrade the top-1 accuracy of the target network so that it is close to a random guess.

In our evaluations, only attacks that degrade the top-1 accuracy to be below the desired value are recorded.

\noindent
\textbf{Targeted Bit-Flip Attack (T-BFA)}~\cite{rakin2021t}: There are three attack types proposed in this work: \textit{N-to-1} attack misclassifies inputs from \textit{N} source classes to 1 target class; \textit{1-to-1} attack misclassifies inputs from 1 source class to 1 target class; and \textit{1-to-1 stealthy} attack does the same as the previous, but aims to keep the other class classification function the same.
The N-to-1 attack is successful if all inputs from the N source classes are misclassified to the target class.
In case of 1-to-1 and 1-to-1 stealthy attacks, the images from the source class are divided into two sets.
One is used in the attack to find the vulnerable bits.
The other set is reserved for calculating the \textit{attack success rate (ASR)} -- which is the percentage of images in the set that are misclassified to the target class.
In the original evaluation setting~\cite{rakin2021t}, the authors terminate the attack if the ASR reaches at least 99.99\% for both 1-to-1 and 1-to-1 stealthy attacks.
Furthermore, for the 1-to-1 stealthy attack, the attack also stops in case the ASR does not change for three consecutive iterations.
For our evaluations, we kept the same stopping criteria for the attacks.

In our evaluations, for the N-to-1 attack, only successful attacks that misclassify all images from the N classes to the 1 target class are studied.
In terms of 1-to-1 attack, since the attack stops only after ASR reaches at least $99.99\%$ and the number of images in each class of our used datasets is not more than $1000$, all attacks reach $100\%$ ASR.
As for 1-to-1 stealthy attacks, because of the stopping criteria, some attacks do not achieve $100\%$ ASR.\\

\noindent
\textbf{Targeted Attack with Limited Bit Flips (TA-LBF)}~\cite{bai2021targeted}: The goal is to misclassify one given sample to an attacker's chosen target class while ensuring stealthiness, by utilizing binary integer programming. 
Selecting the target weight bits is determined by an optimization approach rather than using a heuristic strategy.
An attack is successful if the given sample is misclassified to the chosen target class.

In our evaluations, only successful TA-LBF attacks are analyzed.
We would like to note that in the case of the TA-LBF attack on VGG-11 trained for the CIFAR-100 dataset, we only obtained one successful attack among $200$ attacks.
Thus, we have omitted the results for this setting.

Two datasets were used for the evaluations: CIFAR-10 and CIFAR-100.
The datasets were created using subsets of the tiny images dataset~\cite{torralba200880} with more clear labels.
The details of the creation of both datasets can be found in~\cite{krizhevsky2009learning}.

\noindent
\textbf{CIFAR-10} consists of $60,000$ $32\times32$ pixels color images in $10$ classes with $6000$ images per class.
The ten classes are airplane, automobile, bird, cat, deer, dog, frog, horse, ship, and truck.
The training set contains $50,000$ images and the remaining $10,000$ images are for testing.
There are $1000$ images from each class in the test set.

\noindent
\textbf{CIFAR-100} also consists of $60,000$ $32\times32$ pixels color images.
But there are $100$ classes, each containing $600$ images.
The classes in CIFAR-100 are mutually exclusive with that in CIFAR-10.
For example, CIFAR-100 contains the class pickup truck, but pictures of pickup trucks do not appear in CIFAR-10.
The test dataset contains $10,000$ images, $100$ per class.

We note that following the original attack settings, our experiments only focus on the test images of the datasets.

\noindent
\textbf{Neural network architectures}.
For CIFAR-10 dataset, we have evaluated the attacks against ResNet-20 and VGG-11~\cite{DBLP:journals/corr/SimonyanZ14a}.
And for CIFAR-100 dataset, we have analyzed ResNet-18 and VGG-11.
The quantization method is described in subsection~\ref{sec:quantization}.

\subsection{Evaluation Details}
\label{sec:evalsettings}
Our evaluation repeats each experiment $100$ times for each attack method and each dataset.
The specific settings are described below.

\noindent
\textbf{CIFAR-10}:

\noindent
BFA: Each experiment carries out one attack.
Following the setting in~\cite{rakin2019bit}, the attack aims to degrade the top-1 accuracy below 11\%.
We record the number of bit-flips for successful attacks that achieve this goal.

\noindent
T-BFA: Each experiment corresponds to one attack.
    \begin{itemize}
        \item \textit{N-to-1}: Each experiment randomly selects one class as the target class and $1,000$ images from all classes.
        The goal of the attack is to (mis)classify those $1,000$ images to the target class.
        The number of bit-flips for attacks that reach this goal is recorded.
        \item \textit{1-to-1}: Each experiment randomly generates one target class, one source class, and $500$ images from the source class.
        The goal of the attack is to misclassify all those $500$ images from the source class to the target class.
        According to our discussion of ASR earlier, in this case, ASR is the percentage of misclassified inputs from the remaining $500$ images in the source class.
        The number of bit flips for each attack that reaches ASR at least $99.99\%$ (effectively $100\%$) are recorded.
        \item \textit{1-to-1 stealthy}: Each experiment randomly generates one target class, one source class, $500$ images from the source class, and $500$ images from any other class apart from the source class (those images will be used for maintaining stealthiness).
        In this case, ASR is the percentage of input misclassified to the target class from the remaining $500$ images in the source class.
        The number of bit flips is recorded if the ASR reaches at least $99.99\%$ (effectively $100\%$) or does not change for three consecutive iterations.
    \end{itemize}

\noindent
TA-LBF: Each experiment computes the attack $10$ times, each with one of the $10$ target classes and one randomly generated image from a different class to be misclassified to this target class.
The number of bit flips are recorded for only successful attacks.

\noindent
\textbf{CIFAR-100}:

\noindent
BFA: Each experiment carries out one attack, aiming to degrade the top-1 accuracy below $1.2\%$.
We record the number of bit-flips for successful attacks that achieve this goal.\\

\noindent
T-BFA: Each experiment corresponds to one attack.
    \begin{itemize}
        \item \textit{N-to-1}: Each experiment randomly selects one class as the target class, and $1,000$ images from all classes.
        The goal of the attack is to (mis)classify those $1,000$ images to the target class.
        The number of bit-flips for attacks that reach this goal is recorded.
        \item \textit{1-to-1}: Each experiment randomly generates one target class, one source class, and $50$ images from the source class.
        The goal of the attack is to misclassify all those $50$ images from the source class to the target class.
        In this case, ASR is the percentage of misclassified inputs from the remaining $50$ images in the source class.
        The number of bit flips for each attack that reaches ASR at least $99.99\%$ (effectively $100\%$) are recorded.
        \item \textit{1-to-1 stealthy}: Each experiment randomly generates one target class, one source class, $50$ images from the source class, and $50$ images from any other class apart from the source class (those images will be used for maintaining stealthiness).
        ASR is the percentage of input misclassified to the target class from the remaining $50$ images in the source class.
        The number of bit flips is recorded if the ASR reaches at least $99.99\%$ (effectively $100\%$) or does not change for three consecutive iterations.
    \end{itemize}

\noindent
TA-LBF: Each experiment randomly selects one class as the target class and executes one attack.
The attack goal is to misclassify one randomly generated image (from a class different from the target class) to the target class.
The number of bit flips is recorded for successful attacks that achieve this goal.

\noindent
\textbf{Attacks results on protected implementation.}
The number of bit flips for each attack on the protected implementation can be computed using the changed bits from attacks on the unprotected implementation directly.

In particular, for each attack, we first record the changes in all the weight values.
For example, suppose that after one attack on a $4-$bit quantized neural network, two weights with the original value of $-1$ are changed to $7$, and one weight with the original value of $-2$ is changed to $6$.
In total, we need $3$ bit flips. 

Then, with this information, we can find the Hamming distance (i.e., the number of bit flips needed) between those weight values using our binary code.
For example, with $C_{(7.3)}$, according to Table~\ref{tab:hamming}, changing the codeword corresponding to $-1$ to that corresponding to $7$ needs $7$ bit flips.
Changing the codeword corresponding to $-2$ to codeword for $6$ needs $7$ bit flips.
In total, we need $21$ bit flips.

\noindent
\textbf{Attack success rate for T-BFA 1-to-1 stealthy attack.}
As mentioned before, due to the attack stop criteria, the attack success rates vary for different experiments in the case of T-BFA 1-to-1 stealthy attack.
A summary of the attack success rates for this attack can be found in Table~\ref{tab:asr-1-1-s}.

\begin{table}[tb]
\centering
\begin{tabular}{|c|c|c|c|c|c|}
\hline
Dataset & Model & Quantization & avg & min & max \\ \hline
\multirow{4}{*}{CIFAR-10} & \multirow{2}{*}{ResNet-20} & 4-bit & $97.50$ & $93.00$ & $99.40$ \\ \cline{3-6} 
 &  & 8-bit & $97.51$ & $94.80$ & $99.80$ \\ \cline{2-6} 
 & \multirow{2}{*}{VGG-11} & 4-bit & $46.86$ & $0.00$ & $98.60$ \\ \cline{3-6} 
 &  & 8-bit & $5.83$ & $0.00$ & $89.60$ \\ \hline
\multirow{4}{*}{CIFAR-100} & \multirow{2}{*}{ResNet-18} & 4-bit & $8.66$ & $0.00$ & $100$ \\ \cline{3-6} 
 &  & 8-bit & $5.72$ & $0.00$ & $100$ \\ \cline{2-6} 
 & \multirow{2}{*}{VGG-11} & 4-bit & $0.96$ & $0.00$ & $28.00$ \\ \cline{3-6} 
 &  & 8-bit & $0.52$ & $0.00$ & $12.00$ \\ \hline
\end{tabular}
\vspace{3mm}
\caption{Attack success rates of T-BFA 1-to-1 stealthy attacks.}
\label{tab:asr-1-1-s}
\end{table}

\subsection{Fault Attack Results}
\label{sec:results}

\begin{table*}[tb]
    \centering
    \scalebox{1.01}{
    \begin{NiceTabular}{c|c|c|c|c|c|c|c|c|c|c|c|c|c|c|}\cline{4-15}
        \multicolumn{3}{c|}{} & \multicolumn{12}{c|}{\# of bit flips} \\\cline{4-15} 
        \multicolumn{3}{c|}{} & \multicolumn{3}{c|}{Original} & \multicolumn{3}{c|}{$C_{(7,3)}$} & \multicolumn{3}{c|}{$C_{(8,4)}$} & \multicolumn{3}{c|}{$C_{(9,4)}$} \\\hline
        \multicolumn{1}{|c|}{Dataset} & Attack & Model & min & avg & max & min & avg & max & min & avg & max & min & avg & max \\\hline
        \rowcolor{\rowcone}\cellcolor{white} \Block[draw]{10-1}{\rotate CIFAR-10}& \cellcolor{white} & ResNet-20   & 4 & 14.87 & 52 & 28 & 100.15 & 331 & 32 & 114.32 & 380 & 32 & 115.02 & 383 \\
        \rowcolor{\rowctwo}\cellcolor{white}& \cellcolor{white}\multirow{-2}{*}{BFA} & VGG-11  & 36 & 79.83 & 200 & 233 & 450.4 & 1068 & 260 & 499.72 & 1180 & 271 & 525.81 & 1246 \\\cline{2-15}
        \rowcolor{\rowcone}\cellcolor{white} & \cellcolor{white} & ResNet-20  & 9 & 19.86 & 29 & 63 & 138.57 & 200 & 72 & 158.28 & 228 & 72 & 158.43 & 229 \\
        \rowcolor{\rowctwo}\cellcolor{white} & \cellcolor{white}\multirow{-2}{*}{\makecell{\cellcolor{white}T-BFA\\\cellcolor{white} \textit{N-to-1}}}   & VGG-11  & 34 & 49.79 & 376 & 235 & 329.93 & 2493 & 268 & 375.64 & 2828 & 269 & 379.62 & 2865 \\\cline{2-15}
        \rowcolor{\rowcone}\cellcolor{white} & \cellcolor{white} & ResNet-20 & 6 & 12.02 & 19 & 42 & 83.79 & 133 & 48 & 95.72 & 152 & 48 & 95.81 & 152 \\
         \rowcolor{\rowctwo}\cellcolor{white}& \cellcolor{white}\multirow{-2}{*}{\makecell{\cellcolor{white}T-BFA\\\cellcolor{white} \textit{1-to-1}}}   & VGG-11 & 30 & 51.41 & 96 & 182 & 324.28 & 575 & 208 & 367.72 & 648 & 212 & 375.45 & 669 \\\cline{2-15}
        \rowcolor{\rowcone}\cellcolor{white}& \cellcolor{white} & ResNet-20 & 7 & 15.29 & 35 & 49 & 106.94 & 245 & 56 & 122.2 & 280 & 56 & 122.23 & 280 \\
        \rowcolor{\rowctwo}\cellcolor{white} & \cellcolor{white}\multirow{-2}{*}{\makecell{\cellcolor{white}T-BFA\\\cellcolor{white} \textit{1-to-1 stealthy}}}   & VGG-11 & 3 & 12.37 & 37 & 7 & 82.33 & 250 & 8 & 94.04 & 284 & 10 & 94.68 & 287 \\\cline{2-15}
        \rowcolor{\rowcone}\cellcolor{white} & \cellcolor{white} & ResNet-20 & 3 & 5.23 & 11 & 21 & 36.33 & 77 & 24 & 41.47 & 88 & 24 & 41.56 & 88 \\
        \rowcolor{\rowctwo}\cellcolor{white}&  \cellcolor{white}\multirow{-2}{*}{TA-LBF}  & VGG-11 & 4 & 11 & 43 & 17 & 73.54 & 301 & 20 & 83.91 & 344 & 21 & 84.53 & 344 \\\hline\hline
       \rowcolor{\rowcone}\cellcolor{white} \Block[draw]{10-1}{\rotate CIFAR-100}& \cellcolor{white} & ResNet-18 & 11 & 171.05 & 614 & 77 & 879.17 & 2466 & 88 & 968.96 & 2664 & 88 & 1021.32 & 2848 \\
       \rowcolor{\rowctwo}\cellcolor{white}& \cellcolor{white}\multirow{-2}{*}{BFA} & VGG-11 & 47 & 498.83 & 1281 & 309 & 1828.19 & 4311 & 352 & 1958.24 & 4568 & 356 & 2138.04 & 5060 \\\cline{2-15}
       \rowcolor{\rowcone}\cellcolor{white}& \cellcolor{white} & ResNet-18 & 19 & 68.66 & 111 & 133 & 467.07 & 746 & 152 & 531.28 & 848 & 152 & 535.67 & 856 \\
       \rowcolor{\rowctwo}\cellcolor{white} &\cellcolor{white} \multirow{-2}{*}{\makecell{\cellcolor{white}T-BFA\\\cellcolor{white} \textit{N-to-1}}}   & VGG-11 & 42 & 178.61 & 508 & 258 & 1179.32 & 3342 & 288 & 1335.28 & 3784 & 300 & 1357.31 & 3842 \\\cline{2-15}
       \rowcolor{\rowcone}\cellcolor{white}& \cellcolor{white}& ResNet-18 & 15 & 32.28 & 123 & 99 & 218.64 & 840 & 112 & 248.48 & 956 & 114 & 250.92 & 963 \\
       \rowcolor{\rowctwo}\cellcolor{white}&\cellcolor{white}\multirow{-2}{*}{\makecell{\cellcolor{white}T-BFA\\ \cellcolor{white}\textit{1-to-1}}}     & VGG-11 & 31 & 173.04 & 1050 & 214 & 1152.15 & 6559 & 244 & 1306.6 & 7384 & 245 & 1324.21 & 7575 \\\cline{2-15}
      \rowcolor{\rowcone}\cellcolor{white}& \cellcolor{white} & ResNet-18  & 3 & 4.65 & 25 & 18 & 32.22 & 172 & 20 & 36.76 & 196 & 21 & 36.87 & 197 \\
        \rowcolor{\rowctwo}\cellcolor{white}& \cellcolor{white}\multirow{-2}{*}{\makecell{\cellcolor{white}T-BFA\\\cellcolor{white} \textit{1-to-1 stealthy}}}   & VGG-11 & 3 & 3.47 & 10 & 21 & 24.29 & 70 & 24 & 27.76 & 80 & 24 & 27.76 & 80 \\\cline{2-15}
       \rowcolor{\rowcone}\cellcolor{white}& \cellcolor{white} & ResNet-18 & 7 & 30.51 & 78 & 49 & 209.76 & 531 & 56 & 239 & 604 & 56 & 240.27 & 609 \\
       \rowcolor{\rowctwo}\cellcolor{white}& \cellcolor{white}\multirow{-2}{*}{TA-LBF} & VGG-11  & 52 & 148.50 & 198 & 364 & 1032 & 1386 & 416 & 1178 & 1584 & 416 & 1180.50 & 1584\\\hline                                                     
    \end{NiceTabular}}
    \vspace{3mm}
    \caption{Number of bit flips required for each attack on $4-$bit quantized networks with the original two's complement encoding and with our chosen binary codes.}
    \label{tab:4-bit-result}
\end{table*}

\begin{table*}
    \centering
    \scalebox{1.01}{
    \begin{NiceTabular}{c|c|c|c|c|c|c|c|c|c|c|c|c|c|c|}
    % [code-before = 
    % \rowcolors{1}{\rowcone}{\rowctwo}]
    \cline{4-15}
    \multicolumn{3}{c|}{} &
    \Block{1-12}{\# of bit flips}
    % \multicolumn{12}{c|}{\# of bit flips} 
    \\\cline{4-15} 
    \multicolumn{3}{c|}{} & \multicolumn{3}{c|}{Original} & \multicolumn{3}{c|}{$C_{(12,3)}$} & \multicolumn{3}{c|}{$C_{(13,4)}$} & \multicolumn{3}{c|}{$C_{(14,4)}$} \\\hline
    \multicolumn{1}{|c|}{Dataset} & Attack & Model & min & avg & max & min & avg & max & min & avg & max & min & avg & max \\\hline
    \rowcolor{\rowcone}\cellcolor{white}\Block[draw]{10-1}{\rotate CIFAR-10}&\cellcolor{white} & ResNet-20  & 8 & 19 & 57 & 96 & 222.72 & 660 & 96 & 223.34 & 662 & 104 & 241.96 & 717 \\
    \rowcolor{\rowctwo}\cellcolor{white}& \cellcolor{white} \multirow{-2}{*}{BFA}  & VGG-11  & 49 & 105.73 & 200 & 564 & 1041.59 & 1919 & 578 & 1070.86 & 1988 & 627 & 1161.97 & 2162 \\\cline{2-15}
    \rowcolor{\rowcone}\cellcolor{white} & \cellcolor{white} & ResNet-20  & 12 & 21.9 & 29 & 144 & 262.53 & 348 & 144 & 262.62 & 348 & 156 & 284.52 & 377 \\
    \rowcolor{\rowctwo}\cellcolor{white} & \cellcolor{white}\multirow{-2}{*}{\makecell{\cellcolor{white}T-BFA\\ \cellcolor{white}\textit{N-to-1}}}  & VGG-11 & 47 & 81.75 & 207 & 471 & 931.38 & 2415 & 484 & 939.34 & 2432 & 525 & 1018.23 & 2637 \\\cline{2-15}
    \rowcolor{\rowcone}\cellcolor{white} & \cellcolor{white} & ResNet-20 & 8 & 13.77 & 22 & 96 & 165.09 & 264 & 96 & 165.14 & 264 & 104 & 178.91 & 286 \\
    \rowcolor{\rowctwo}\cellcolor{white} & \cellcolor{white}\multirow{-2}{*}{\makecell{\cellcolor{white}T-BFA\\ \cellcolor{white}\textit{1-to-1}}}   & VGG-11 & 42 & 83.29 & 216 & 462 & 934.85 & 2559 & 466 & 943.04 & 2564 & 504 & 1021.89 & 2778 \\\cline{2-15}
    \rowcolor{\rowcone}\cellcolor{white}&\cellcolor{white}  & ResNet-20 & 8 & 16.34 & 37 & 96 & 195.81 & 444 & 96 & 195.84 & 444 & 104 & 212.16 & 481 \\
    \rowcolor{\rowctwo}\cellcolor{white}& \cellcolor{white} \multirow{-2}{*}{\makecell{\cellcolor{white}T-BFA\\ \cellcolor{white}\textit{1-to-1 stealthy}}}& VGG-11 & 3 & 5.6 & 53 & 12 & 64.68 & 609 & 14 & 64.86 & 610 & 15 & 70.24 & 661 \\\cline{2-15}
    \rowcolor{\rowcone}\cellcolor{white}& \cellcolor{white} & ResNet-20 & 2 & 6.24 & 12 & 24 & 74.18 & 144 & 24 & 74.25 & 144 & 26 & 80.44 & 156 \\
    \rowcolor{\rowctwo}\cellcolor{white}& \cellcolor{white}\multirow{-2}{*}{TA-LBF}& VGG-11 & 4 & 26.21 & 81 & 36 & 301.54 & 942 & 38 & 302.78 & 946 & 41 & 327.96 & 1025 \\\hline\hline
    \rowcolor{\rowcone}\cellcolor{white} \Block[draw]{9-1}{\rotate CIFAR-100}& \cellcolor{white} & ResNet-18 & 17 & 104.47 & 360 & 198 & 1040.09 & 3333 & 200 & 1065.94 & 3448 & 217 & 1156.85 & 3746 \\
   \rowcolor{\rowctwo}\cellcolor{white}&    \cellcolor{white}\multirow{-2}{*}{BFA}& VGG-11 & 60 & 428.05 & 1106 & 678 & 3206.22 & 6765 & 688 & 3328 & 7070 & 746 & 3602.61 & 7658 \\\cline{2-15}
   \rowcolor{\rowcone}\cellcolor{white}& \cellcolor{white} & ResNet-18 & 21 & 34.99 & 82 & 246 & 413.13 & 978 & 248 & 415.26 & 980 & 269 & 450.21 & 1062 \\
   \rowcolor{\rowctwo}\cellcolor{white}&   \cellcolor{white}\multirow{-2}{*}{\makecell{\cellcolor{white}T-BFA\\ \cellcolor{white}\textit{N-to-1}}}& VGG-11 & 42 & 214.96 & 796 & 483 & 2501.43 & 9213 & 490 & 2520.7 & 9286 & 532 & 2733.58 & 10070 \\\cline{2-15}
   \rowcolor{\rowcone}\cellcolor{white}&\cellcolor{white} & ResNet-18 & 14 & 33.16 & 348 & 168 & 396.42 & 4173 & 168 & 396.92 & 4174 & 182 & 430.08 & 4522 \\
    \rowcolor{\rowctwo}\cellcolor{white}& \cellcolor{white} \multirow{-2}{*}{\makecell{\cellcolor{white}T-BFA\\ \cellcolor{white}\textit{1-to-1}}}   & VGG-11 & 29 & 150.13 & 758 & 348 & 1767.55 & 8823 & 348 & 1776.08 & 8894 & 377 & 1925.37 & 9646 \\\cline{2-15}
    \rowcolor{\rowcone}\cellcolor{white}& \cellcolor{white} & ResNet-18 & 3 & 3.99 & 21 & 36 & 47.73 & 249 & 36 & 47.78 & 250 & 39 & 51.77 & 271 \\
    \rowcolor{\rowctwo}\cellcolor{white}&  \cellcolor{white}\multirow{-2}{*}{\makecell{\cellcolor{white}T-BFA\\ \cellcolor{white}\textit{1-to-1 stealthy}}}& VGG-11 & 3 & 3.19 & 6 & 36 & 38.28 & 72 & 36 & 38.28 & 72 & 39 & 41.47 & 78 \\\cline{2-15}
    \rowcolor{\rowcone}\cellcolor{white}& \cellcolor{white}\multirow{1}{*}{TA-LBF} & ResNet-18 & 9 & 46.63 & 112 & 108 & 557.01 & 1341 & 108 & 557.86 & 1342 & 117 & 604.49 & 1454 \\\hline
                                                                                                 % &    & VGG-11 & & & & & & & & & & & & \\\hline                                                                                       
    \end{NiceTabular}}
    \vspace{3mm}
    \caption{Number of bit flips required for each attack on $8-$bit quantized networks with the original two's complement encoding and with our chosen binary codes.}
    \label{tab:8-bit-results}
\end{table*}

\begin{figure}[tb]
    \centering
    \includegraphics[width=0.48\textwidth]{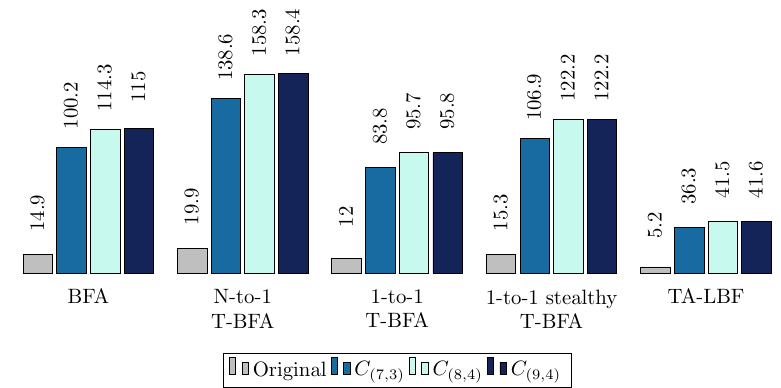}
    {\footnotesize(a) 4-bit ResNet-20 on CIFAR-10}
    
    \vspace{0.5cm}
    \includegraphics[width=0.48\textwidth]{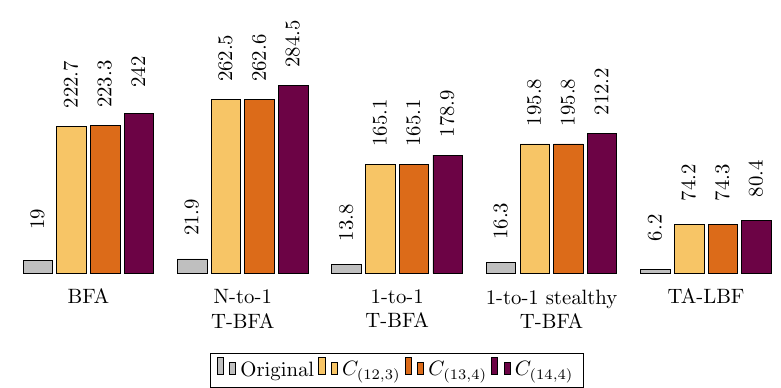}
    {\footnotesize(b) 8-bit ResNet-20 on CIFAR-10}
    \caption{Averaged number of bit flips required for BFA, T-BFA, and TA-LBF attacks on quantized ResNet-20, with and without \method countermeasure.}
    \label{fig:bar-chart}
\end{figure}

The minimum, average, and maximum number of bit flips required for the attacks described in Section~\ref{sec:attacksetting} are listed in Table~\ref{tab:4-bit-result} ($4-$bit quantized networks) and Table~\ref{tab:8-bit-results} ($8-$bit quantized networks).

As discussed in Section~\ref{sec:bitflip-statistics}, in the case of $4-$bit quantized networks, we expect to see around $6\times$  ($11\times$ in the case of $8-$bit quantized networks) increase in the number of bit flips required for protected implementations as compared to unprotected implementations.
The results for T-BFA and TA-LBF in Tables~\ref{tab:4-bit-result} and~\ref{tab:8-bit-results} agree with our expectations.

For example, TA-LBF attacks on the 4-bit quantized VGG-11 need on average $10.999$ bit flips.
When $C_{(9,4)}$ is used for the protection, the average number of bit flips is $84.531$, which is $7.69\times$ more compared to the unprotected implementation.
As shown in Table~\ref{tab:linear94}, using $C_{(9,4)}$, codewords corresponding to integers that only differ in the MSB have Hamming distances $8$.
This explains why the increase in the number of bit flips required.

In summary, using $C_{(7,3)}$ for the countermeasure, for T-BFA N-to-1 attacks, the average number of bit flips for protected implementations is between $6.60\times$ (VGG-11, CIFAR-100) to $6.98\times$ (ResNet-20, CIFAR-10) of that for unprotected implementations.
Moreover, we have $6.31\times$ (VGG-11, CIFAR-10) to $6.97\times$ (ResNet-20, CIFAR-10) for T-BFA 1-to-1 attacks; $6.66\times$ (VGG-11, CIFAR-10) to $7\times$ (VGG-11, CIFAR-100) for 1-to-1 stealthy attacks; and $6.69\times$ (VGG-11, CIFAR-10) to $6.95\times$ (VGG-11, CIFAR-100) for TA-LBF attacks.

When $C_{(8,4)}$ is used as our binary code, the change in the averaged number of bit flips for protected implementation compared to that for unprotected implementation is $7.48\times$ (VGG-11, CIFAR-100) to $7.97\times$ (ResNet-20, CIFAR-10) in the case of T-BFA N-to-1 attacks, $7.15\times$ (VGG-11, CIFAR-10) to $7.96\times$ (ResNet-20, CIFAR-10) in T-BFA 1-to-1 attacks, $7.60\times$ (VGG-11, CIFAR-10) to $8\times$ (VGG-11, CIFAR-100) in T-BFA 1-to-1 stealthy attacks, and $7.63\times$ (VGG-11, CIFAR-10) to $7.93\times$ (VGG-11, CIFAR-100) in TA-LBF attacks.

As for $C_{(9,4)}$, the protection provided compared to the original two's complement implementation is $7.6\times$ (VGG-11, CIFAR-100) to $7.98\times$ (ResNet-20, CIFAR-10) for T-BFA N-to-1 attacks, $7.3\times$ (VGG-11, CIFAR-10) to $7.97\times$ (ResNet-20, CIFAR-10) for T-BFA 1-to-1 attacks, $7.65\times$ (VGG-11, CIFAR-10) to $8\times$ (VGG-11, CIFAR-100) for T-BFA 1-to-1 stealthy attacks, and $7.69\times$ (VGG-11, CIFAR-10) to $7.95\times$ (VGG-11, CIFAR-100) for TA-LBF attacks.

However, for protection against BFA attacks, the increase in the average number of bit flips fluctuates more.
Using $C_{(7,4)}$, the minimum change is $3.66\times$ (VGG-11, CIFAR-100) and the maximum change is $6.74\times$ (ResNet-20, CIFAR-10).
With $C_{(8,4)}$, the range is between $3.93\times$ (VGG-11, CIFAR-100) to $7.69\times$ (ResNet-20, CIFAR-10).
In case of $C_{(9,4)}$ we have $4.29\times$ (VGG-11, CIFAR-100) to $7.74\times$ (ResNet-20, CIFAR-10).

To explain this observation, we take a further analysis of the bit flip attacks on the weight values by each attack method.
There were in total $61,812$ weight value changes achieved by BFA attacks, among which $20.18\%$ resulted in more than $1-$bit changes.
However, for T-BFA and TA-LBF attacks, $2-4$ bit changes sum up to between $0.44\%$ (T-BFA N-to-1) and $0.85\%$ (TA-LBF).
For example, with the code $C_{(7,4)}$, the distances between the codewords are at most $7$.
Compared to $2-$, $3-$, or $4-$bit flips, $7$ is just $3.5\times$, $2.3\times$, or $1.75\times$ increase.
This explains why the protection against BFA is not as strong as that against other attacks.

In the case of $8-$bit quantized neural networks, we have similar results.
The change in the averaged number of bit flips ranges from $7.49\times$ (BFA attack on VGG-11 for CIFAR-100 with $C_{(12,3)}$ for the countermeasure) to $13\times$ (T0BFA 1-to-1 stealthy attack on VGG-11 for CIFAR-100 with $C_{(12,3)}$ for the countermeasure).

We note that for some cases, the number of bit flips is smaller for T-BFA 1-to-1 stealthy attack than for T-BFA 1-to-1 attack, e.g. for $4-$bit quantized VGG-11.
This is because of the stopping criteria of the attacks -- for 1-to-1 attacks, we reach ASR $100\%$ while for 1-to-1 stealthy attacks, we may not reach ASR $100\%$ (see Table~\ref{tab:asr-1-1-s}).

For a better overview, we include two bar charts demonstrating the averaged number of bit flips required for BFA, T-BFA, and TA-LBF attacks on quantized ResNet-20 in Figure~\ref{fig:bar-chart}, where Figures~\ref{fig:bar-chart} (a) and (b) detail the results for 4-bit and 8-bit quantized ResNet-20, respectively.
Apart from the relative numbers, it is also important to look at the total amount of bit flips, which becomes impractical for the attacker if a detection function is run periodically on top of \method.
The shortest time required for a successful Rowhammer attack is over $11$ seconds for a $4-$bit quantization and $23$ seconds for an $8-$bit one, thus providing a large time window to run a notification/recovery procedure.
We discuss this in more detail in Section~\ref{sec:speeds}.

\section{Discussion}
\label{sec:discussion}
\subsection{Overheads}

\noindent
\textbf{Memory overhead.} 
The memory overhead is directly related to the type of code used.
We state the details for each code in Tables~\ref{tab:overhead_4bit} and \ref{tab:overhead_8bit} for $4-$bit and $8-$bit networks, respectively.
Generally, overheads for $8-$bit quantization are lower in relative numbers, while providing better detection properties.

Another look at the memory overhead would be from the perspective of how many times more bit flips the attacker needs for an additional bit of overhead.
If we take the averaged results from Figure~\ref{fig:overview}, this ratio is as follows: $C_{(7,3)}: 2.2\times$, $C_{(8,4)}: 1.875\times$, $C_{(9,4)}: 1.52\times$, $C_{(12,3)}: 2.825\times$, $C_{(13,4)}: 2.28\times$, $C_{(14,4)}: 2.07\times$.
We can see that the best trade-offs are provided by the shortest codes -- $C_{(7,3)}$ and $C_{(12,3)}$, as adding additional bits does not greatly increase the number of required bit flips.
Naturally, the decision needs to be made based on some risk analysis, depending on the attacker's assumed capabilities and the criticality of the deployed model.

\noindent
\textbf{Computation time overhead.}
The computation time overhead is related to the encoding and decoding functions.
As the encoding is done only once before the model is deployed in the memory, the overhead is negligible compared to the time the model stays in the memory.
The decoding is done on-demand (e.g. before every few inference executions).
It depends on the security settings on how often we fetch the encoded values from the memory and decode them.
In terms of relative time compared to the inference, on our setup (Intel Core i5-11400F @ 2.6GHz, 64 GB of RAM @ 3.2 GHz), it took on average $2.7\times$ the decoding (e.g. $0.00083$s for $4-$bit ResNet-20 compared to $0.00031$s for decoding $C_{(7,3)}$).
Therefore, even for a single inference, the time overhead is around $37.3\%$, but we argue that it is not necessary to calculate this for every single inference as attacking several weights with a Rowhammer takes a much longer time. 
According to~\cite{kwong2020rambleed}, where the Rowhammer attack was used for a recovery of RSA keys, the average flipping speed was $\approx 0.31$ bits/second.
For example, let us take the worst case scenario from Figure~\ref{fig:bar-chart} (a): TA-LBF attack on $4-$bit quantized ResNet-20 protected with the shortest code -- $C_{(7,3)}$.
The attacker would require more than $11$ seconds in an ideal scenario to mount a successful attack\footnote{To compare, without \method, the attacker would only need $\approx\! 1.6$ seconds to cause enough bit flips in the same setting.}. 
Therefore, if the inference engine is running constantly, it would be sufficient to run the decoding routine every $1000$ inference executions, making the time overhead negligible.

\label{sec:overhead}
\begin{table}[tb]
\centering
\begin{tabular}{|c|c|c|c|c|}\cline{3-5}
\multicolumn{2}{c}{} & \multicolumn{3}{|c|}{Time overhead (s)} \\\hline 
Dataset                    & Model      & $C_{(7,3)}$ & $C_{(8,4)}$ & $C_{(9,4)}$ \\ \hline
\multirow{2}{*}{CIFAR-10}  & ResNet-20  & 0.00031 &  0.00032  &  0.00032  \\ \cline{2-5} 
                           & VGG-11     & 0.01147  &  0.01147  & 0.01147   \\ \hline \hline
\multirow{2}{*}{CIFAR-100} & ResNet-18  & 0.01350 &  0.01350  &  0.01310  \\ \cline{2-5} 
                           & VGG-11     & 0.03340 &  0.03341  &  0.03341 \\ \cline{2-5} 
                           % & WRN-28-4   & 0.34515 &  0.34961  &  0.40679  \\ \cline{2-5} 
                           % & WRN-28-8   & 1.40590 &  1.40317  &  1.64091 \\ 
                           \hline\hline
\multicolumn{2}{|c|}{Memory overhead} & $75\%$ & $100\%$ & $125\%$\\\hline
\end{tabular}
\vspace{3mm}
\caption{Time and memory overheads for decoding weights for quantized networks with $4-$bit precision.}
\label{tab:overhead_4bit}
\end{table}

\begin{table}[tb]
\centering
\begin{tabular}{|c|c|c|c|c|}\cline{3-5}
\multicolumn{2}{c}{} & \multicolumn{3}{|c|}{Time overhead (s)} \\\hline 
Dataset                    & Model      & $C_{(12,3)}$ & $C_{(13,4)}$ & $C_{(14,4)}$ \\ \hline
\multirow{2}{*}{CIFAR-10}  & ResNet-20  & 0.00032 &  0.00032  &  0.00033  \\ \cline{2-5} 
                           & VGG-11     & 0.01146  &  0.01147  & 0.01146   \\ \hline \hline
\multirow{2}{*}{CIFAR-100} & ResNet-18  & 0.01351 &  0.01350  &  0.01351  \\ \cline{2-5} 
                           & VGG-11     &  0.03341 & 0.03341   & 0.03340 \\ \cline{2-5} 
                           % & WRN-28-4   & 034515 &  034961  &  040679  \\ \cline{2-5} 
                           % & WRN-28-8   & 140590 &  140317  &  164091 \\ 
                           \hline\hline
\multicolumn{2}{|c|}{Memory overhead} & $50\%$ & $62.5\%$ & $75\%$\\\hline
\end{tabular}
\vspace{3mm}
\caption{Time and memory overheads for decoding weights for quantized networks with $8-$bit precision.}
\label{tab:overhead_8bit}
\end{table}

\subsection{\method with Detection Capability}
\label{sec:speeds}
There is a possibility to utilize \method as a detection countermeasure, similarly to~\cite{li2021radar,javaheripi2021hashtag}.
As the decoding of weights takes a much smaller amount of time compared to a successful Rowhammer attack, and the number of bit flips required with \method protection increases significantly, it is possible to choose a setting so that weight corruption gets detected before a successful attack happens.
Therefore, we can set a detection alarm that runs the decoding function every few seconds irrespective of whether the inference is running or not to provide us with a sufficient security margin.
In this scenario, considering the Rowhammer speed is relatively stable with a given DRAM frequency, we could detect the attack attempt with a $100\%$ detection rate.

\subsection{Choice of Code}
From the attack results, we can see that when the attack mostly changes just $1$ bit of the weight values, with a proper encoding design, we can increase the number of bit flips by a factor close to the maximum distance achieved by codewords of the code.
In case the attack requires more bit flips in a single weight value, longer codes, and therefore larger distances between codewords, will be needed for similar protection levels.
As the time overhead does not change significantly with longer codes, the only thing to consider in this trade-off scenario would be the memory overhead.

\section{Conclusion}
\label{sec:conclusion}
In this paper, we presented \method, an encoding-based countermeasure against bit-flip attacks on neural networks.
The protection is based on Hamming codes and extended Hamming codes with various parameters, providing high detection rates for adversarial (but also non-adversarial) bit flips.
We focused on protecting quantized neural networks with $4-$bit and $8-$bit quantization as these are the most common targets of these attacks in the literature.
Through extensive experimental evaluation, we demonstrated the utility of our countermeasure on state-of-the-art bit flip attacks and multiple publicly available models and datasets.
Time and memory overheads of \method show a good trade-off between the resources and security.

% A potential but challenging follow-up would be to implement the inference computation on encoded data, allowing utilization of side-channel resistant codes and thus extending the protection to these attacks.

% \begin{acks}
% \end{acks}

%%
\newpage
\bibliographystyle{IEEEtran}
\bibliography{biblio}

\begin{thebibliography}{10}
\providecommand{\url}[1]{#1}
\csname url@samestyle\endcsname
\providecommand{\newblock}{\relax}
\providecommand{\bibinfo}[2]{#2}
\providecommand{\BIBentrySTDinterwordspacing}{\spaceskip=0pt\relax}
\providecommand{\BIBentryALTinterwordstretchfactor}{4}
\providecommand{\BIBentryALTinterwordspacing}{\spaceskip=\fontdimen2\font plus
\BIBentryALTinterwordstretchfactor\fontdimen3\font minus
  \fontdimen4\font\relax}
\providecommand{\BIBforeignlanguage}[2]{{%
\expandafter\ifx\csname l@#1\endcsname\relax
\typeout{** WARNING: IEEEtran.bst: No hyphenation pattern has been}%
\typeout{** loaded for the language `#1'. Using the pattern for}%
\typeout{** the default language instead.}%
\else
\language=\csname l@#1\endcsname
\fi
#2}}
\providecommand{\BIBdecl}{\relax}
\BIBdecl

\bibitem{szegedy2013intriguing}
C.~Szegedy, W.~Zaremba, I.~Sutskever, J.~Bruna, D.~Erhan, I.~Goodfellow, and
  R.~Fergus, ``Intriguing properties of neural networks,'' \emph{arXiv preprint
  arXiv:1312.6199}, 2013.

\bibitem{biggio2018wild}
B.~Biggio and F.~Roli, ``Wild patterns: Ten years after the rise of adversarial
  machine learning,'' in \emph{Proceedings of the 2018 ACM SIGSAC Conference on
  Computer and Communications Security}, 2018, pp. 2154--2156.

\bibitem{batina2022implementation}
L.~Batina, S.~Bhasin, J.~Breier, X.~Hou, and D.~Jap, ``On implementation-level
  security of edge-based machine learning models,'' in \emph{Security and
  Artificial Intelligence: A Crossdisciplinary Approach}.\hskip 1em plus 0.5em
  minus 0.4em\relax Springer, 2022, pp. 335--359.

\bibitem{mittal2021survey}
S.~Mittal, H.~Gupta, and S.~Srivastava, ``A survey on hardware security of dnn
  models and accelerators,'' \emph{Journal of Systems Architecture}, vol. 117,
  p. 102163, 2021.

\bibitem{breier2018practical}
J.~Breier, X.~Hou, D.~Jap, L.~Ma, S.~Bhasin, and Y.~Liu, ``Practical fault
  attack on deep neural networks,'' in \emph{Proceedings of the 2018 ACM SIGSAC
  Conference on Computer and Communications Security}.\hskip 1em plus 0.5em
  minus 0.4em\relax ACM, 2018, pp. 2204--2206.

\bibitem{hong2019terminal}
S.~Hong, P.~Frigo, Y.~Kaya, C.~Giuffrida, and T.~Dumitraș, ``Terminal brain
  damage: Exposing the graceless degradation in deep neural networks under
  hardware fault attacks,'' in \emph{28th USENIX Security Symposium (USENIX
  Security 19)}, 2019, pp. 497--514.

\bibitem{liu2017fault}
Y.~Liu, L.~Wei, B.~Luo, and Q.~Xu, ``Fault injection attack on deep neural
  network,'' in \emph{2017 IEEE/ACM International Conference on Computer-Aided
  Design (ICCAD)}.\hskip 1em plus 0.5em minus 0.4em\relax IEEE, 2017, pp.
  131--138.

\bibitem{breier2021sniff}
J.~Breier, D.~Jap, X.~Hou, S.~Bhasin, and Y.~Liu, ``Sniff: reverse engineering
  of neural networks with fault attacks,'' \emph{IEEE Transactions on
  Reliability}, vol.~71, no.~4, pp. 1527--1539, 2021.

\bibitem{rakin2020tbt}
A.~S. Rakin, Z.~He, and D.~Fan, ``Tbt: Targeted neural network attack with bit
  trojan,'' in \emph{Proceedings of the IEEE/CVF Conference on Computer Vision
  and Pattern Recognition}, 2020, pp. 13\,198--13\,207.

\bibitem{breier2022foobar}
J.~Breier, X.~Hou, M.~Ochoa, and J.~Solano, ``Foobar: Fault fooling backdoor
  attack on neural network training,'' \emph{IEEE Transactions on Dependable
  and Secure Computing}, vol.~20, no.~3, pp. 1895--1908, 2023.

\bibitem{rakin2021ra}
A.~S. Rakin, L.~Yang, J.~Li, F.~Yao, C.~Chakrabarti, Y.~Cao, J.-s. Seo, and
  D.~Fan, ``{RA-BNN: Constructing robust \& accurate binary neural network to
  simultaneously defend adversarial bit-flip attack and improve accuracy},''
  \emph{arXiv preprint arXiv:2103.13813}, 2021.

\bibitem{he2020defending}
Z.~He, A.~S. Rakin, J.~Li, C.~Chakrabarti, and D.~Fan, ``Defending and
  harnessing the bit-flip based adversarial weight attack,'' in
  \emph{Proceedings of the IEEE/CVF Conference on Computer Vision and Pattern
  Recognition}, 2020, pp. 14\,095--14\,103.

\bibitem{kim2014flipping}
Y.~Kim, R.~Daly, J.~Kim, C.~Fallin, J.~H. Lee, D.~Lee, C.~Wilkerson, K.~Lai,
  and O.~Mutlu, ``Flipping bits in memory without accessing them: An
  experimental study of dram disturbance errors,'' \emph{ACM SIGARCH Computer
  Architecture News}, vol.~42, no.~3, pp. 361--372, 2014.

\bibitem{rakin2019bit}
A.~S. Rakin, Z.~He, and D.~Fan, ``Bit-flip attack: Crushing neural network with
  progressive bit search,'' in \emph{Proceedings of the IEEE/CVF International
  Conference on Computer Vision}, 2019, pp. 1211--1220.

\bibitem{rakin2021t}
A.~S. Rakin, Z.~He, J.~Li, F.~Yao, C.~Chakrabarti, and D.~Fan, ``T-bfa:
  Targeted bit-flip adversarial weight attack,'' \emph{IEEE Transactions on
  Pattern Analysis and Machine Intelligence}, vol.~44, no.~11, pp. 7928--7939,
  2021.

\bibitem{bai2021targeted}
J.~Bai, B.~Wu, Y.~Zhang, Y.~Li, Z.~Li, and S.-T. Xia, ``Targeted attack against
  deep neural networks via flipping limited weight bits,'' \emph{arXiv preprint
  arXiv:2102.10496}, 2021.

\bibitem{lecun2015deep}
Y.~LeCun, Y.~Bengio, and G.~Hinton, ``Deep learning,'' \emph{Nature}, vol. 521,
  no. 7553, pp. 436--444, 2015.

\bibitem{hubara2017quantized}
I.~Hubara, M.~Courbariaux, D.~Soudry, R.~El-Yaniv, and Y.~Bengio, ``Quantized
  neural networks: Training neural networks with low precision weights and
  activations,'' \emph{The Journal of Machine Learning Research}, vol.~18,
  no.~1, pp. 6869--6898, 2017.

\bibitem{nagel2021white}
M.~Nagel, M.~Fournarakis, R.~A. Amjad, Y.~Bondarenko, M.~Van~Baalen, and
  T.~Blankevoort, ``A white paper on neural network quantization,'' \emph{arXiv
  preprint arXiv:2106.08295}, 2021.

\bibitem{breier2022practical}
J.~Breier and X.~Hou, ``How practical are fault injection attacks, really?''
  \emph{IEEE Access}, vol.~10, pp. 113\,122--113\,130, 2022.

\bibitem{boneh2001importance}
D.~Boneh, R.~A. DeMillo, and R.~J. Lipton, ``On the importance of checking
  cryptographic protocols for faults,'' in \emph{Advances in
  Cryptology—EUROCRYPT’97: International Conference on the Theory and
  Application of Cryptographic Techniques Konstanz, Germany, May 11--15, 1997
  Proceedings}.\hskip 1em plus 0.5em minus 0.4em\relax Springer, 1997, pp.
  37--51.

\bibitem{baksi2022survey}
A.~Baksi, S.~Bhasin, J.~Breier, D.~Jap, and D.~Saha, ``A survey on fault
  attacks on symmetric key cryptosystems,'' \emph{ACM Computing Surveys},
  vol.~55, no.~4, pp. 1--34, 2022.

\bibitem{mutlu2019rowhammer}
O.~Mutlu and J.~S. Kim, ``Rowhammer: A retrospective,'' \emph{IEEE Transactions
  on Computer-Aided Design of Integrated Circuits and Systems}, vol.~39, no.~8,
  pp. 1555--1571, 2019.

\bibitem{hou2021physical}
X.~Hou, J.~Breier, D.~Jap, L.~Ma, S.~Bhasin, and Y.~Liu, ``Physical security of
  deep learning on edge devices: Comprehensive evaluation of fault injection
  attack vectors,'' \emph{Microelectronics Reliability}, vol. 120, p. 114116,
  2021.

\bibitem{luo2021deepstrike}
Y.~Luo, C.~Gongye, Y.~Fei, and X.~Xu, ``Deepstrike: Remotely-guided fault
  injection attacks on dnn accelerator in cloud-fpga,'' in \emph{2021 58th
  ACM/IEEE Design Automation Conference (DAC)}.\hskip 1em plus 0.5em minus
  0.4em\relax IEEE, 2021, pp. 295--300.

\bibitem{chen2021proflip}
H.~Chen, C.~Fu, J.~Zhao, and F.~Koushanfar, ``Proflip: Targeted trojan attack
  with progressive bit flips,'' in \emph{Proceedings of the IEEE/CVF
  International Conference on Computer Vision}, 2021, pp. 7718--7727.

\bibitem{selmke2016attack}
B.~Selmke, J.~Heyszl, and G.~Sigl, ``Attack on a dfa protected aes by
  simultaneous laser fault injections,'' in \emph{2016 Workshop on Fault
  Diagnosis and Tolerance in Cryptography (FDTC)}.\hskip 1em plus 0.5em minus
  0.4em\relax IEEE, 2016, pp. 36--46.

\bibitem{patrick2017lightweight}
C.~Patrick, B.~Yuce, N.~F. Ghalaty, and P.~Schaumont, ``Lightweight fault
  attack resistance in software using intra-instruction redundancy,'' in
  \emph{Selected Areas in Cryptography--SAC 2016: 23rd International
  Conference, St. John's, NL, Canada, August 10-12, 2016, Revised Selected
  Papers 23}.\hskip 1em plus 0.5em minus 0.4em\relax Springer, 2017, pp.
  231--244.

\bibitem{zussa2014efficiency}
L.~Zussa, A.~Dehbaoui, K.~Tobich, J.-M. Dutertre, P.~Maurine,
  L.~Guillaume-Sage, J.~Clediere, and A.~Tria, ``Efficiency of a glitch
  detector against electromagnetic fault injection,'' in \emph{2014 Design,
  Automation \& Test in Europe Conference \& Exhibition (DATE)}.\hskip 1em plus
  0.5em minus 0.4em\relax IEEE, 2014, pp. 1--6.

\bibitem{he2017fpga}
W.~He, J.~Breier, S.~Bhasin, N.~Miura, and M.~Nagata, ``An fpga-compatible
  pll-based sensor against fault injection attack,'' in \emph{2017 22nd Asia
  and South Pacific Design Automation Conference (ASP-DAC)}.\hskip 1em plus
  0.5em minus 0.4em\relax IEEE, 2017, pp. 39--40.

\bibitem{patranabis2015fault}
S.~Patranabis, A.~Chakraborty, and D.~Mukhopadhyay, ``Fault tolerant infective
  countermeasure for aes,'' in \emph{Security, Privacy, and Applied
  Cryptography Engineering: 5th International Conference, SPACE 2015, Jaipur,
  India, October 3-7, 2015, Proceedings 5}.\hskip 1em plus 0.5em minus
  0.4em\relax Springer, 2015, pp. 190--209.

\bibitem{mennink2020beyond}
B.~Mennink, ``Beyond birthday bound secure fresh rekeying: Application to
  authenticated encryption,'' in \emph{Advances in Cryptology--ASIACRYPT 2020:
  26th International Conference on the Theory and Application of Cryptology and
  Information Security, Daejeon, South Korea, December 7--11, 2020,
  Proceedings, Part I 26}.\hskip 1em plus 0.5em minus 0.4em\relax Springer,
  2020, pp. 630--661.

\bibitem{baksi2021default}
A.~Baksi, S.~Bhasin, J.~Breier, M.~Khairallah, T.~Peyrin, S.~Sarkar, and S.~M.
  Sim, ``Default: Cipher level resistance against differential fault attack,''
  in \emph{Advances in Cryptology--ASIACRYPT 2021: 27th International
  Conference on the Theory and Application of Cryptology and Information
  Security, Singapore, December 6--10, 2021, Proceedings, Part II 27}.\hskip
  1em plus 0.5em minus 0.4em\relax Springer, 2021, pp. 124--156.

\bibitem{breier2017feeding}
J.~Breier and X.~Hou, ``Feeding two cats with one bowl: On designing a fault
  and side-channel resistant software encoding scheme,'' in \emph{Topics in
  Cryptology--CT-RSA 2017: The Cryptographers’ Track at the RSA Conference
  2017, San Francisco, CA, USA, February 14--17, 2017, Proceedings}.\hskip 1em
  plus 0.5em minus 0.4em\relax Springer, 2017, pp. 77--94.

\bibitem{breier2019evaluating}
J.~Breier, X.~Hou, and Y.~Liu, ``On evaluating fault resilient encoding schemes
  in software,'' \emph{IEEE Transactions on Dependable and Secure Computing},
  vol.~18, no.~3, pp. 1065--1079, 2019.

\bibitem{guo2013recomputing}
X.~Guo and R.~Karri, ``Recomputing with permuted operands: A concurrent error
  detection approach,'' \emph{IEEE Transactions on Computer-Aided Design of
  Integrated Circuits and Systems}, vol.~32, no.~10, pp. 1595--1608, 2013.

\bibitem{li2021radar}
J.~Li, A.~S. Rakin, Z.~He, D.~Fan, and C.~Chakrabarti, ``Radar: Run-time
  adversarial weight attack detection and accuracy recovery,'' in \emph{2021
  Design, Automation \& Test in Europe Conference \& Exhibition (DATE)}.\hskip
  1em plus 0.5em minus 0.4em\relax IEEE, 2021, pp. 790--795.

\bibitem{javaheripi2021hashtag}
M.~Javaheripi and F.~Koushanfar, ``Hashtag: Hash signatures for online
  detection of fault-injection attacks on deep neural networks,'' in \emph{2021
  IEEE/ACM International Conference On Computer Aided Design (ICCAD)}.\hskip
  1em plus 0.5em minus 0.4em\relax IEEE, 2021, pp. 1--9.

\bibitem{wang2023aegis}
J.~Wang, Z.~Zhang, M.~Wang, H.~Qiu, T.~Zhang, Q.~Li, Z.~Li, T.~Wei, and
  C.~Zhang, ``Aegis: Mitigating targeted bit-flip attacks against deep neural
  networks,'' in \emph{32nd USENIX Security Symposium (USENIX Security 23)},
  2023, pp. 2329--2346.

\bibitem{liu2023neuropots}
Q.~Liu, J.~Yin, W.~Wen, C.~Yang, and S.~Sha, ``$\{$NeuroPots$\}$: Realtime
  proactive defense against $\{$Bit-Flip$\}$ attacks in neural networks,'' in
  \emph{32nd USENIX Security Symposium (USENIX Security 23)}, 2023, pp.
  6347--6364.

\bibitem{ozdenizci2022improving}
O.~{\"O}zdenizci and R.~Legenstein, ``Improving robustness against stealthy
  weight bit-flip attacks by output code matching,'' in \emph{Proceedings of
  the IEEE/CVF Conference on Computer Vision and Pattern Recognition}, 2022,
  pp. 13\,388--13\,397.

\bibitem{zhou2023dnn}
R.~Zhou, S.~Ahmed, A.~S. Rakin, and S.~Angizi, ``{DNN}-defender: An in-{DRAM}
  deep neural network defense mechanism for adversarial weight attack,''
  \emph{arXiv preprint arXiv:2305.08034}, 2023.

\bibitem{macwilliams+sloane}
F.~J. MacWilliams and N.~J.~A. Sloane, \emph{The Theory of Error-Correcting
  Codes}.\hskip 1em plus 0.5em minus 0.4em\relax North-Holland Publishing
  Company, 1977.

\bibitem{ling2004coding}
S.~Ling and C.~Xing, \emph{Coding Theory: A First Course}.\hskip 1em plus 0.5em
  minus 0.4em\relax Cambridge University Press, 2004.

\bibitem{torralba200880}
A.~Torralba, R.~Fergus, and W.~T. Freeman, ``80 million tiny images: A large
  data set for nonparametric object and scene recognition,'' \emph{IEEE
  Transactions on Pattern Analysis and Machine Intelligence}, vol.~30, no.~11,
  pp. 1958--1970, 2008.

\bibitem{krizhevsky2009learning}
A.~Krizhevsky, G.~Hinton \emph{et~al.}, ``Learning multiple layers of features
  from tiny images,'' 2009.

\bibitem{DBLP:journals/corr/SimonyanZ14a}
\BIBentryALTinterwordspacing
K.~Simonyan and A.~Zisserman, ``Very deep convolutional networks for
  large-scale image recognition,'' in \emph{3rd International Conference on
  Learning Representations, {ICLR} 2015, San Diego, CA, USA, May 7-9, 2015,
  Conference Track Proceedings}, Y.~Bengio and Y.~LeCun, Eds., 2015. [Online].
  Available: \url{http://arxiv.org/abs/1409.1556}
\BIBentrySTDinterwordspacing

\bibitem{kwong2020rambleed}
A.~Kwong, D.~Genkin, D.~Gruss, and Y.~Yarom, ``Rambleed: Reading bits in memory
  without accessing them,'' in \emph{41st {IEEE} Symposium on Security and
  Privacy (S\&P)}, 2020.

\end{thebibliography}

% \clearpage
% \appendix

% \input{appendix}

\end{document}